\documentclass[aip,jcp,reprint]{revtex4-2}
\usepackage{amsmath,amssymb}
\usepackage{dcolumn}% Align table columns on decimal point
\usepackage{bm}
\usepackage[version=4]{mhchem}
\usepackage{chemscheme}
\usepackage{soul}
\usepackage{chemfig}
\usepackage{chemformula}
\usepackage{verbatim} 
\usepackage{braket}
\usepackage{xcolor}
\usepackage{graphicx}
\usepackage[symbol]{footmisc}
\usepackage{floatrow}
\usepackage{xr}
\usepackage{hyperref}
\newcommand{\revS}[1]{{\color{black}{#1}}}
\newcommand{\revX}[1]{{\color{black}{#1}}}

\begin{document}

\title{Evidence of orbital mixing upon ionization via Cooper minimum photoelectron dynamics in epichlorohydrin.
Experiment and Theory.} 

\author{L. Schio}
\affiliation{CNR - Istituto Officina dei Materiali (IOM), Laboratorio TASC, Area Science Park Basovizza, Trieste, Italy}
\affiliation{Department of Basic and Applied Science for Engineering (SBAI), Sapienza University, Rome, Italy}
\author{M. Alagia}
\affiliation{CNR - Istituto Officina dei Materiali (IOM), Laboratorio TASC, Area Science Park Basovizza, Trieste, Italy}
\author{T. Moitra}
\affiliation{DTU Chemistry, Technical University of Denmark, Kemitorvet 207, 2800 Kongens Lyngby, Denmark}
\affiliation{Department of Physical and Theoretical Chemistry, Faculty of Natural Sciences, Comenius University, SK-84215, Bratislava, Slovakia}
\author{D. Toffoli}
\affiliation{Dipartimento di Scienze Chimiche e Farmaceutiche, Universit{\`a} degli Studi di Trieste, Trieste, Italy}
\author{A. Ponzi}
\affiliation{Institut 
Ruder Bo{\v s}kovi{\'c}, Bijeni{\v s}ka cesta 54, 10000 Zagreb, Croatia}
\author{M. Stener} 
\affiliation{Dipartimento di Scienze Chimiche e Farmaceutiche, Universit{\`a} degli Studi di Trieste, Trieste, Italy}
\author{S. Coriani}
\affiliation{DTU Chemistry, Technical University of Denmark, Kemitorvet 207, 2800 Kongens Lyngby, Denmark}
\author{P. Decleva}
\email{decleva@units.it}
\affiliation{Dipartimento di Scienze Chimiche e Farmaceutiche, Universit{\`a} degli Studi di Trieste, Trieste, Italy}
\author{O. Rebrov}  
\affiliation{Department of Physics, University of Stockholm, Stockholm, Sweden}
\author{V. Zhaunerchyk}  
\affiliation{Department of Physics, University of Gothenburg, Gothenburg, Sweden}
\author{M. Larsson} 
\affiliation{Department of Physics, University of Stockholm, Stockholm, Sweden}
\author{S. Falcinelli} 
\affiliation{Dipartimento di Ingegneria Civile ed Ambientale, Universit{\`a} degli Studi di Perugia, Perugia, Italy}
\author{A. A. Dias} 
\affiliation{Laboratory for Instrumentation, Biomedical Engineering, and Radiation Physics,
Departamento de Fisica, Facultade de Ciencias e Tecnologia,
Universidade NOVA de Lisboa, Capacarica, Portugal}
\author{D. Catone} 
\affiliation{CNR-ISM, Area della Ricerca di Roma Tor Vergata, Via del Fosso del Cavaliere 100, Roma, Italy}
\author{S. Turchini} 
\affiliation{CNR-ISM, Area della Ricerca di Roma Tor Vergata, Via del Fosso del Cavaliere 100, Roma, Italy}
\author{N. Zema}  
\author{F. Salvador} 
\affiliation{CNR-ISM, Area della Ricerca di Roma Tor Vergata, Via del Fosso del Cavaliere 100, Roma, Italy}
\author{D. Benedetti}  
\affiliation{CNR - Istituto Officina dei Materiali (IOM), Laboratorio TASC, Area Science Park Basovizza, Trieste, Italy}
\author{D. Vivoda} 
\affiliation{Elettra-Sincrotrone Trieste, Area Science Park, Basovizza, Trieste, Italy}
\author{B. Botta}  
\affiliation{Department of Chemistry and Technologies of Drugs, Sapienza University, I-00183, Rome, Italy}
\author{S. Stranges}
\email{stefano.stranges@uniroma1.it}
\affiliation{Department of Chemistry and Technologies of Drugs, Sapienza University, I-00183, Rome, Italy} 
\affiliation{CNR - Istituto Officina dei Materiali (IOM), Laboratorio TASC, Area Science Park Basovizza, Trieste, Italy}

\date{\today}

% \begin{tocentry}
% \includegraphics[width=1\textwidth]{Figures/TOC_Epi_FINALE_4x8cm_970DPI.jpg}
% %Some journals require a graphical entry for the Table of Contents.
% %This should be laid out ``print ready'' so that the sizing of the
% %text is correct.

% %Inside the \texttt{tocentry} environment, the font used is Helvetica
% %8\,pt, as required by \emph{Journal of the American Chemical
% %Society}.

% %The surrounding frame is 9\,cm by 3.5\,cm, which is the maximum
% %permitted for  \emph{Journal of the American Chemical Society}
% %graphical table of content entries. The box will not resize if the
% %content is too big: instead it will overflow the edge of the box.

% %This box and the associated title will always be printed on a
% %separate page at the end of the document.
% \end{tocentry}

\begin{abstract}

% \revS{Can we try and sharpen the abstract a bit? Be more explicit about being the first to present experimental evidence of orbital mixing?}

A peculiar electron correlation effect, leading to orbital rotation upon ionization, theoretically predicted long ago, was never experimentally characterized. The effect is expected to appear prominently in the photoionization of chiral molecules, due to the lack of symmetry constraints to wave-functions mixing. This is observed to have a profound effect on the photoelectron dynamics, as here demonstrated by investigating $\beta$ asymmetry parameters and partial cross-section observables in the Cl 3p Cooper minimum region of epichlorohydrin, a chiral prototype system.

Angle-resolved photoelectron spectroscopy with tunable synchrotron radiation allowed measuring Cooper minimum $\beta$ oscillations, which were observed for solely two valence photoionization channels. 
% The nature and number of channels exhibiting such dynamical behavior, and the extent of the observed oscillation amplitudes, could not be explained by predictions based on HF and DFT.
The nature and number of channels exhibiting such dynamical behavior, along with the extent of the observed oscillation amplitudes, could not be accounted for by predictions based on Hartree-Fock (HF) and Density Functional Theory (DFT). 
These features could only be explained by incorporating correlation effects, which mix single-hole configurations of identical symmetry, in the characterization of the four lowest-lying molecular cation states, via equation-of-motion coupled cluster singles and doubles Dyson orbitals.
% but only considering 
% correlation effects, 
% which mix different single-hole configurations, of same symmetry, in characterizing the four lowest lying molecular cation states
% HF and DFT. The experimental findings were only realized considering correlation effects, which mix different single-hole configurations of same symmetry, hence enabling characterization of the four lowest lying molecular cation states.
\end{abstract}

\maketitle

\section{Introduction}
Removal of a single electron (for instance, by photoionization) is the most elementary excitation of a many-body system. In the Independent Particle Approximation (IPA), it is directly connected with the negative orbital energies, and photoionization 
has provided 
an enormous understanding 
of the orbital structure of molecules. 
%Actually, IPA is only a first-order model. 
However, the IPA is only a first-order model and thus offers only a simplified representation of electron interactions.
Correlation effects alter this picture in several ways, so photoionization \revS{offers} a unique laboratory to study correlation effects.
% Indeed, a very strong interplay between photoionization experiments and electronic structure theory has started since the very beginning of the new technique.
Indeed, a strong interplay between photoionization experiments and electronic structure theory has existed since the inception of the technique.~\cite{Vilesov:1961:EnergyDO,
Siegbahn:1969:ESCA,
Turner:1962:jcp.1962} 
%Nordling:1958:PhysRev.105.1676,

Among the several effects displayed and analyzed, one of the most elusive is {\em orbital mixing}, i.e., the fact that the ``electron-hole'' orbital relative to a particular final ionic state, instead of resembling one of the occupied Hartree-Fock (HF) orbitals of the ground state, is 
% a significant linear combination of two, or more of them, 
a linear combination, with significantly large coefficients, of two or more of them, 
but still of norm close to one. 
In more precise terms, the Dyson 
orbital,~\cite{Pickup:1977:Dyson,Arneberg:1982:Dyson,Oana:2007,Ortiz:2020:Dyson} 
which characterizes ionization to a specific final state, \revS{becomes} a mixture of ground-state orbitals and can be considered as a kind of orbital rotation upon ionization.

Although discovered theoretically many years ago,~\cite{VonNiessen:1982:hole-mixing,Kuleff:2014} orbital mixing has remained (\revS{until now}) a 
%\revS{purely} 
theoretical curiosity, since it cannot be assessed on pure energy grounds--that is, by means of photoelectron spectroscopy. 
%integrate some of referee 1's comments....\\
% This article provides a combined experimental/theoretical investigation of an orbital blending
% that can be manifest in the photoelectron angular distributions from a small molecule, epichlorohydrin. 
% It is well established that electron correlation effects in molecular ionizations in
% the inner valence region lead to a breakdown of the independent electron approximation, with shake up and shake down processes. Experimentally these are readily detected as satellite
% transitions with energies displaced from the main line. 
% However, with the advent of more
% advanced computational approaches (e.g., EOM-CCSD) 
% it is becoming apparent that a weaker
% correlation effects can mix the canonical orbitals in the outer valence , where traditionally
% independent electron approximation for ionization has been assumed to be valid. 
% %The authors identify this as an “orbital rotation”. 
% A difficulty for experimental verification is that the eigenvalues are
% little perturbed by this, so the experimental shifts in ionization energy go undetected.
%
\revS{Specifically, 
it is well established that
electron‑correlation effects in inner‑valence molecular ionization lead to a breakdown of the independent‑electron approximation, giving rise to shake‑up and shake‑down processes. Experimentally, these are readily observed as satellite transitions at energies displaced from the main line. However, with the advent of more advanced computational methods, 
%(e.g., EOM‑CCSD), 
it has become clear that weaker correlation effects can also mix canonical orbitals in the outer‑valence region, where the independent‑electron picture has traditionally been assumed to hold. A challenge for experimental verification is that such mixing only minimally perturbs the eigenvalues, causing any shifts in ionization energies to remain essentially undetectable.}
Only transition properties, typically dynamical photoionization observables like cross sections or angular distributions, can probe the nature of the Dyson orbital and reveal differences in its composition. 
This effect is characteristic of the outer valence shell, where electron correlation is generally weaker, hence not to the point of complete breakdown of the single-particle picture, as is commonly observed in the inner valence region.~\cite{Cederbaum:1978:ionis,
Cederbaum:1986:correlation}
% The effect is typical of the outer valence shell, where correlation effects are generally less intense, hence not to the point of a complete breakdown of the single particle picture (which instead characterizes the inner valence region).~\cite{Cederbaum:1978:ionis,
% Cederbaum:1986:correlation}
It may be associated with differential relaxation properties of the HF orbitals, e.g., due to different localization/delocalization structure, or differential correlation of their main atomic orbital (AO) composition. This is often the case between s, p and d AO’s in transition metal complexes,~\cite{Rohmer:1973} and even between s, p orbitals relative to different shells, like in elements of the second and third period. 
Obviously, the mixing is easier when initial orbital energies are quite close, so that a relatively small perturbation can induce significant mixing. 
Thus, mixing is expected to be common in molecules with a dense manifold of outer valence orbitals of the same symmetry, which are the only ones allowed to mix. 
This is commonly the case in low-symmetry systems, and indeed orbital mixing may prove rather frequent in chiral molecules with no symmetry.     
A recent surge of interest in the phenomenon has been stimulated by the possibility of inducing the formation of a 
non-stationary electron wavepacket by photoionization and studying attosecond phenomena,\cite{Kuleff:2014} for instance charge migration in pump-probe experiments, e.g, in propiolic acid~\cite{Despre:2018}  and the PENNA molecule.~\cite{Mignolet:2014} 
An example has also been detected in studying photoelectron circular dichroism (PECD) in ionization of a chiral molecule.~\cite{Waters:2022:cphc}
% \revS{IT WAS SUGGESTED TO ADD REF TO REF.~\citenum{Wanie:2024}. BUT IS IT RELEVANT?}
% \revP{I think it is not, delete}
%new
Photoelectron dynamics involves the study of properties of photoelectrons emitted from matter.  In the case of randomly-oriented species and in the electric dipole approximation, it entails the knowledge of the partial, angle-integrated, photoionization cross sections ($\sigma$), and of the angular distribution (the $\beta$ asymmetry or anisotropy parameter) as a function of photon energy. Moreover, specific of chiral systems with circularly polarized light, a further angular (dichroic) parameter, $\beta_1$ or $D$, can be obtained, the so-called PECD effect. The experimental determination of these observables is ideally performed with synchrotron radiation (SR), an easily tunable photon source with controlled polarization, and Angle-Resolved PhotoElectron Spectroscopy (ARPES).
One of the most striking and informative dynamical effects exhibited by $\sigma(h\nu)$ and $\beta(h\nu)$ is the Cooper minimum (CM).\cite{Cooper:1962,Manson:1968} This phenomenon manifests itself in the form of a minimum of $\sigma(h\nu)$ when the photoionization process involves electron ejection from an atomic orbital with a radial node, since the minimum occurs because of the change in sign of radial matrix elements governing the transition moment, as the photon energy is increased from the ionization threshold. Concomitant with this $\sigma$ behavior, the $\beta$ parameter is strongly affected and displays, in the same CM energy region, a rapid oscillation characterized by a maximum and minimum. This latter, $\beta_{\textrm{min}}$, occurs at a photon energy which approaches closely that of $\sigma_{\textrm{min}}$. The CM phenomenon was thoroughly investigated theoretically~\cite{Cooper:1962,Manson:1968} and experimentally~\cite{Houlgate:1974,Houlgate:1976,Miller:1977,Torop:1976} as a purely atomic effect. Interestingly, this effect was discovered to persist also in molecular photoionization by Carlson and 
co-workers.~\cite{Carlson:1982:CCl4,
Carlson:1982:CS2-COS,
Carlson:1983:Cl2,
Carlson:1983:HCl,
Carlson:1986:SiCl4,
Carlson:1986} 
Minima in both $\sigma$ and $\beta$ were experimentally observed, and theoretically confirmed, for all cases in which the photoionization process, in the one-electron picture, involves the electron ejection from a non-bonding MO (lone-pair) with strong atomic character due to an AO with radial nodes, namely the S 3p or Cl 3p lone-pairs in the investigated systems. The good agreement between the measured and the calculated profiles of both $\sigma$ and $\beta$ allowed to associate the atomic CM phenomenon with a molecular CM. This concept was further enriched by observing CM dynamical effects, although of lesser extent, in photoionization processes involving $\sigma$-bonding MOs with partial contribution from AOs with radial nodes.~\cite{Carlson:1982:CCl4,Carlson:1982:CS2-COS,Carlson:1983:HCl,Carlson:1983:Cl2,Carlson:1986:SiCl4,Carlson:1986,Potts:1985,Thiel:1984}  Generally, the CM dynamical features of $\beta$ are found to be much more pronounced than those of $\sigma$, making $\beta$ the most sensitive probe with regard to CM dynamics. Such values are easier to measure experimentally than the corresponding $\sigma$ value changes. It was found that the lower the $\beta_{\textrm{min}}$ value, the closer is the nature of the MO to the AO from which it derives, as revealed, for instance, by the Cl 3p AO contributions to different MOs involved in a large number of chlorinated molecules.~\cite{Carlson:1982:CCl4,Carlson:1983:Cl2,Carlson:1986:SiCl4,Carlson:1983:HCl,Carlson:1986,Potts:1985,Thiel:1984,Ponzi:2014}  
Analogous studies were also accomplished on bromine atom CM in bromobenzene.~\cite{Holland:2000,Powis:2015:BrBz,Powis:2017:BrBz} 

The characterization of orbital nature via detection of the CM offers a remarkable experimental possibility of detecting the orbital rotation effect. Because of the complexity and lack of symmetry of chiral molecules, with an associated manifold of closely lying ionic states, such effects can be expected to be quite common in such systems. 
% Although $\beta_1$ decreases fast with increasing electron kinetic energy and is therefore a less sensitive probe of the CM, it will probably be strongly affected by the orbital rotation.~\cite{Waters:2022:cphc}
% Since theoretical simulations of PECD, like in conventional CD, are almost mandatory to interpret the observed signal, the present study has strong implication for future PECD studies.

Monosubstituted oxiranes (ethylene oxides) have been studied experimentally and theoretically as important model chiral molecules. Particularly, methyloxirane (C$_3$H$_6$O, also known as propylene oxide) has been studied with SR by several authors.~\cite{Garcia:2014,Stranges:2005,Turchini:2004,Alberti:2008}  The interaction of chiral molecules with circularly polarized radiation is considered to be a plausible mechanism connected to the abiotic origin of life. In this context, it is worth mentioning that propylene oxide has recently been detected in space as the first organic chiral molecule.~\cite{McGuire:2016} 
(Chloromethyl)oxirane (epichlorohydrin), derived from methyloxirane by hydrogen substitution with chlorine, has also been investigated with SR.~\cite{Daly:2011,Stranges:2011} 
%end new

In the present work, we report very clear evidence of the phenomenon of orbital mixing 
in the asymmetry parameter $\beta$ (and in the cross section $\sigma$) 
in the photoionization of epichlorohydrin, where the participation of Cl 3p AO’s can be experimentally highlighted by the very strong Cooper minimum effect in the asymmetry parameters. 
A high-level theoretical treatment is provided, showing excellent agreement with the experimental results, indicating the interplay of correlation in the bound states, described by accurate Dyson orbitals obtained by the highly correlated equation-of-motion (EOM) coupled cluster singles and double (CCSD) approach,~\cite{Oana:2007,Oana:2009,Moitra:2020,Moitra:2021} coupled with a time-dependent density functional theory (TDDFT) description of the molecular continuum, which includes interchannel coupling.~\cite{Stener:2005:B-spline-TDDFT-noniter,Moitra:2021,Decleva:2022:Tiresia}

\section{Method}
\label{Methods}

\subsection{Experimental} The experimental PE spectra were measured on the Circularly Polarized Beamline~\cite{Derossi:1995} of ELETTRA (Trieste), by photoionization of epichlorohydrin racemic samples (Aldrich, $\ge$ 99\% purity) \revS{with linearly polarized radiation}.  The PE spectra were recorded in the 13–54 eV photon energy range using the ARPES-TPES end station~\cite{Dyke:1997,Innocenti:2008:jpca,Innocenti:2008:ChemEuJ,SchioAlagia:2020} 
%\revS{(OsO$_4$ cited here)} 
by operating the electron spectrometer in the ARPES mode. Both electron detection angles of 0$^\circ$ and 54.7$^\circ$ (the magic angle) with respect to the polarization plane of the linearly polarized radiation were set. The %anisotropy 
asymmetry
parameters $\beta_i(h\nu)$, characterizing the photoelectron angular distribution, were measured as a function of photon energy as derived from the equation $\beta = (R-1)$,\cite{Cooper:1968, Southworth:1982} with 
$R={A(0^\circ)}/{A(54.7^\circ)}$ being the ratio between the peak areas of the considered PE band obtained at $\theta = 0^\circ$ and $\theta = 54.7^\circ$, respectively.
The vapour, generated from the epichlorohydrin liquid sample, was admitted to the interaction region by an effusive source through a nonmagnetic hypodermic needle. The liquid sample was kept at room temperature (23~$^\circ$C) during the measurements, and the gas rating was adjusted to set a constant pressure of 1.0 $\times$ 10$^{-6}$ mbar in the ionization chamber. This pressure was monitored during the experiment. For signal normalization purpose, the incident photon flux was recorded, along with the photoelectron signal, with two different methods, namely as current of a silicon photodiode (IRD-AXUV-100) set as beam stopper after the interaction region, and as current of the gold-coated refocusing mirror mounted at the entrance of the ionization chamber. 
The measurements, performed in a large photon energy range, required using both the Normal Incidence Monochromator (NIM) and the Spherical Grazing monochromator (SGM) of the beamline. 
The PE spectra were corrected by the spectrometer transmission function, as appropriate, considering the change in electron detection efficiency due to the different value of the electron kinetic energy and the selected constant-pass energy value. For this purpose, PE spectra of two reference gases, He and Ar, were recorded at the magic angle and at different kinetic energies and pass energies (5 eV and 10 eV). Furthermore, the PE spectrum of the first four PE bands of epichlorohydrin was systematically acquired at the photon energies 22.5 eV (NIM range) and 34.5 eV (SGM range) and at 0$^\circ$ detection angle, as reference spectra, to monitor possible small changes in the density of the vapour sample in the ionization region during the measurements.

\begin{comment}
Since the $\beta_i(h\nu)$ parameter is obtained from the peak areas of the considered PE band obtained at 
$\theta = 0^\circ$ and 
$\theta = 54.7^\circ$ {(see Section~\ref{SI-sec:extraction} for details)}, the peak areas at the two different angles were referred to the same photon flux and molecular target density.
\end{comment}

\revS{The $\sigma_i(h\nu)$ and $\beta_i(h\nu)$ values were extracted from the spectra by determining the relative intensities of single PE components via a global multipeak fitting {(see 
Section~S1
%Section~\ref{SI-sec:extraction} 
for details)}. Spectra were first normalized to the photon flux and molecular target density, and corrected for the spectrometer transmission. Backgrounds, modeled as an arctan plus linear term, were obtained from spectra without epichlorohydrin using $\chi^2$ fitting and subtracted, including the onset region. This correction was particularly important at low PE kinetic energies. All bands, being asymmetric with longer high-IE tails, were fit using asymmetric bi-Gaussian functions~\cite{Yu:2010,Buys:1972}  (two half-Gaussians sharing a peak maximum) described by four parameters: peak position (E$_0$), left and right  widths ($\sigma_l$, $\sigma_r$), and scale factor $\delta$, this latter being related to the peak area ($\delta(\sigma_l+\sigma_r)/2$) and the peak height 
($\delta/\sqrt{2\pi}$). The first band (24a MO) position was treated as a free parameter per spectrum and aligned to 
Ref.~\citenum{Stranges:2011}, while other bands’ positions were expressed relative to the HOMO and shared across spectra. For the first four bands, each peak retained individual shapes, whereas the six higher-energy bands shared a common shape (single $\sigma_l$ and $\sigma_r$) across photon energies and angles. Scale factors $\delta_i$ were free for each spectrum. All spectra, measured at different photon energies for both $\theta = 0^\circ$ and $\theta = 54.7^\circ$ detection angles, were simultaneously fit in the global multipeak analysis. }

\revS{This procedure provided a consistent measurement of the PE band relative intensities and enabled the extraction of $\beta_i(h\nu)$ and $\sigma_i(h\nu)$ for the valence photoionization channels of epichlorohydrin as a function of photon energy.}
Errors on the epichlorohydrin $\beta$ values were estimated by measuring, with the same experimental procedure, 
well-known 
%anisotropy 
asymmetry
parameters of reference gases, namely Ar, He and O$_2$, in a wide photon energy range and exhibiting a large variety of $\beta$ values.\cite{Holland:1982:Calibration,Kreile:1980,Dehmer:1975}
%\revS{ global fit procedure inserted}
\subsection{Theoretical}
The differential cross section for photoionization relative to a final ionic state $\Psi^{N-1}_i$ (for randomly oriented molecules and linearly polarized light, in the dipole approximation) is given by the well known formula
\begin{equation}
    \frac{d\sigma_i}{d\theta} = \frac{\sigma_i}{4\pi} [1 + \beta_i P_2(\cos\theta)]
\end{equation}
where $\sigma_i$ is the cross section, 
$\beta_i$ is the asymmetry parameter, and $\theta$ is the angle between the electron's linear momentum and the direction of the light.

Calculating the photoionization parameters $\sigma_i$ and $\beta_i$ requires evaluation of the dipole matrix elements between the initial bound state $\Psi^N_0$ and the final continuum state $\Psi^N_{Ei}$ that describes the bound state $\Psi^{N-1}_i$ and the photoelectron with energy $E$ (all other quantum numbers are omitted for simplicity).
%
% \begin{equation}
%     D^{0i} = \langle{\Psi_0^N}|\vec{d}|{\Psi^N_{Ei}}\rangle~.
%      \label{eq:Dif}
% \end{equation}
%
At the single determinant level (IPA), employing a common set of orbitals $\{\phi_p\}$, describing both bound and continuum states, 
%\begin{multline}    
\begin{equation}
\Psi^N_0 = | \phi_1 \ldots \phi_i \ldots \phi_N \rangle 
%\\
\quad \textrm{and} \quad
\Psi^N_{Ei} = | \phi_1 \ldots \phi_E \ldots \phi_N \rangle 
\end{equation}
%\end{multline} 
we have
\begin{equation}
\label{DipoleMatrixElement}
D^{0i}_{E\alpha} = \langle \Psi^N_{Ei} | D_{\alpha} | \Psi^N_0  \rangle   = \langle \phi_E | d_{\alpha} | \phi_i\rangle  
\end{equation}
where the subscript $\alpha$ indicates the electric dipole cartesian component.
In the static DFT approach  all orbitals are eigenstates of the ground state Kohn-Sham hamiltonian, $h_{\textrm{KS}}$,~\cite{Toffoli:2002:Bspline-multicenter}
\begin{equation}
h_{\textrm{KS}} \phi_p = E_p \phi_p~.
\end{equation}
The one-particle Schr{\"o}dinger equation can be solved accurately both for the bound and continuum states employing a
B-spline basis and a Galerkin approach for the continuum eigenstates.~\cite{Brosolo:1994,Fischer:1990} 
%refs{}

In general, occupied bound DFT orbitals are quite close to the Hartree-Fock ones. However, in critical cases, it is interesting to compare the results obtained using the DFT orbitals with those obtained employing the Hartree-Fock orbitals as initial orbitals
{in the Koopmans' description}. 

In fact, correlation effects in the bound states can be taken accurately into account by employing as initial orbital in Eq.~\eqref{DipoleMatrixElement} the Dyson orbital~\cite{Pickup:1977:Dyson,Arneberg:1982:Dyson,Ortiz:2020:Dyson} relative to the initial $\Psi^N_0$ and final ionic state $\Psi^{N-1}_i$ computed from highly correlated wavefunctions~\cite{Oana:2007,Ponzi:2014,Moitra:2022,Tenorio:2022:Molecules}
\begin{equation}
\phi^\textrm{d}_i = \langle \Psi^{N-1}_i | \hat{\psi}(x) | \Psi^N_0  \rangle = \sum_p \gamma_{ip} \phi_p~;  \ 
\gamma_{ip} =   \langle \Psi^{N-1}_i |  a_p | \Psi^N_0 \rangle 
\end{equation}
where $\hat{\psi}(x)= \sum_q \phi_q(x) a_q$ and $a_p$ are field and orbital annihilator operators, respectively.~\cite{Moitra:2021}
Thus, 
\begin{equation}
    D^{0i}_{E\alpha} = \langle 
    \phi_{E}
     | d_\alpha|
\phi^{\textrm{d}}_{{i}}\rangle~.
    \label{eq-D0i-Dyson}
\end{equation}
Moreover, the squared norm of the Dyson orbital, defined as
\begin{equation}
R_i = || \phi_i^\textrm{d} ||^2 = \sum_p \gamma^2_{ip} 
\end{equation}
can be used as the intensities of the bands to simulate the photoelectron spectrum. $R_i$ 
is known as pole strength, spectral strength, or  spectroscopic factor of the final ionic state
$\Psi^{N-1}_i$. 

In this study, we use Dyson orbitals obtained at the equation of motion coupled cluster singles and doubles (EOM-CCSD) level of theory.~\cite{Oana:2007,Oana:2009,Moitra:2020}
Within EOM-CCSD theory, one has distinct left ($\phi_i^{\textrm{d},L}$) and right ($\phi_i^{\textrm{d},R}$) Dyson orbitals~\cite{Oana:2007,Oana:2009,Moitra:2020} 
\begin{equation}
\phi_i^{\textrm{d},L} = \sum_{p} \gamma^L_{ip}\phi_p 
\quad \quad 
\gamma^L_{ip} \equiv \langle \Psi_{0L}^N  |a_p^\dagger |\Psi^{N-1}_{iR} \rangle
\end{equation}
\begin{equation}
\phi_i^{\textrm{d},R} = \sum_{p} \gamma^R_{ip}\phi_p 
\quad \quad 
\gamma^R_{ip} \equiv \langle \Psi^{N-1}_{iL}  |a_p | 
   \Psi^{N}_{0R} \rangle
\end{equation}
In the case of ionization from the electronic ground state
% \begin{align}
%   \gamma^R_p &= \langle \Phi_0 (1+\Lambda) e^{-(T_1+T_2)}|p^\dagger | R^{IP} e^{(T_1+T_2)}\Phi_0 \rangle, \\
%   \gamma^L_p &= \langle \Phi_0 L^{IP} e^{-(T_1+T_2)}  |p|e^{(T_1+T_2)}\Phi_0 \rangle 
% \end{align}
   \begin{equation}
   \gamma^L_{ip} 
%   \equiv
%   \langle \Psi^{N-1}_L  |a_p | 
%   \Psi^{N}_R \rangle
   =
   \langle 
   {\rm{HF}} | L^{\rm{IP}} %_{p}
   %(1+\hat{\Lambda}) 
   e^{(-T_2-T_1)}
  a_p 
  e^{(T_1+T_2)}|{\rm{HF}}\rangle
   %\Psi^{N} 
   %\rangle =
   \label{left_GS}
\end{equation}
\begin{equation}
   \gamma^R_{ip} 
%   \equiv 
%   \langle \Psi_L^N  |a_p^\dagger |
%   \Psi^{N-1}_R \rangle 
   = 
   \langle {\rm{HF}}|(1+\hat{\Lambda})e^{(-T_2-T_1)}
   a_p^\dagger  
   R^{\rm{IP}} %_{p}
   e^{(T_1+T_2)}|{\rm{HF}}\rangle
   \label{right_GS}
   \end{equation}
where we adopted standard EOM-CCSD notation,~\cite{Stanton:93:EOMCC} 
\begin{equation}
    \hat{\Lambda} =\sum_{\lambda_i=\lambda_1,\lambda_2}\bar{t}_{\lambda_i}\tau^\dagger_{\lambda_i}
    \end{equation}
    \begin{equation}
     R^{\rm{IP}} =
     \sum_{i}r_{i}a_{i}+\frac{1}{2}\sum_{aij}r^{a}_{ij} a^{\dagger}_a a_j a_i~.
\end{equation}
 % As a known effect of inclusion of correlation, the spectral strengths of the primary ionic states are lowered and, at the same time, additional states, i.e. satellite or shake-up states, which are characterized by further electronic excitations, gain intensity.
% As right and left Dyson orbitals are different in EOM-CC,
% \revP{one may ponder how to compute the spectroscopic factor, for instance whether one should take the square norm of just one of the two  Dyson orbitals (left or right),
% or to use the product of their norms (geometric norm), or a third recipe altogether}.~\cite{Vidal_PCCP,Vidal_PCCP_Erratum}
Since right and left Dyson orbitals are different in EOM-CC, we here use as EOM-CC spectroscopic factors the dot product of the left and right Dyson orbital pairs,~\cite{Moitra:2021} 
\begin{equation}
    R_i = \sum_p (\gamma^{{L}}_{ip} \gamma^{{R}}_{ip})~~.
\end{equation}
The individual products $\gamma^{L}_{ip} %\times
\gamma^{R}_{ip}$ will be used as weights for specific $p$ orbital contributions to a given ionisation $i$.

Because of the non-variational nature of CC,
the calculation of the photoelectron
transition dipole matrix element, 
Eq.~\eqref{eq-D0i-Dyson},
also requires generalization. 
As discussed in the SI of ref.~\citenum{Moitra:2020},
the transition strength (matrix element) between initial state $0$ and final state $i$ is written as
(we omit the subscript $E$ for ease of notation)
\begin{align}
S^{0i}_{\alpha\beta} = \frac{1}{2} \left\{D_\alpha^{0i} D_\beta^{i0} + (D_\beta^{0i} D_\alpha^{i0}
)^*\right\} 
\end{align} 
where the left and right moments are
\begin{align}
 D_{\alpha}^{0i} = \langle \Psi_0 |d_\alpha| \Psi_i \rangle^{L}~~; \quad \quad
 D_{\alpha}^{i0} = \langle \Psi_i |d_\alpha| \Psi_0 \rangle^{R}~.
\end{align}
The labels $L$ or $R$ highlight once again that in EOM-CC the above are not
true scalar products, since they are not Hermitian, 
$D_\alpha^{i0} \neq
(D_\alpha^{0i})^*$.
%
%The continuum orbital takes the role of your diffuse gaussian $\phi_d$. Then as I understand one should have
%
% Thus, within EOM-CC formalism, the photoelectron transition moments can be written as,
We then write
\begin{align}
D_{L,\alpha}^{0i} = \langle {\phi^{\textrm{d},L}_{i}} |d_\alpha| \phi_{E} \rangle ~~; 
% \quad \quad 
% \phi^{\textrm{d},L}_{i0}  = \sum_p \gamma^L_p \phi_p
% \nonumber
\quad \quad 
D_{R,\alpha}^{i0}  = \langle \phi_{E} |d_\alpha| \phi^{\textrm{d},R}_{i} \rangle
% \quad \quad 
% \phi^R_{i0}  = \sum_p \gamma^R_p \phi_p
\label{T-dyson-epsilon}
\end{align}
% From now onwards, we will use ${\phi^R_{i0}}$ instead of ${\phi^{\textrm{d},R}_{i0}}$
% and similarly for the left one.
The transition moments in Eq.~\eqref{T-dyson-epsilon}
are now true scalar products
(as computed in our program {Tiresia~\cite{Toffoli:2024:Tiresia}}) between the continuum and
the left and right Dyson orbitals.
Using the above expressions,
we then consider the photoelectron transition strength matrix
\begin{align}
\nonumber
S^{i0}_{\alpha\beta} 
&\equiv
 \langle \Psi_i | d_\alpha | \Psi_0 \rangle \langle \Psi_0 | d_\beta | \Psi_i
 \rangle \\\nonumber
 &= \frac{1}{2} \left[ 
\langle \phi_{E} |d_\alpha| \phi^{\textrm{d},R}_{i} \rangle \langle \phi^{\textrm{d},L}_{i} |d_\beta| \phi_{E}\rangle 
\right.\\\nonumber 
& \;\;\;\;\; \left. 
+ \left( 
\langle \phi_{E} |d_\beta| \phi^{\textrm{d},R}_{i} \rangle \langle
\phi^{\textrm{d},L}_{i} |d_\alpha| \phi_{E} \rangle \right)^*  \right]  \\
 &= \frac{1}{2}  \left[ \langle \phi_{E} |d_\alpha| \phi^{\textrm{d},R}_{i} \rangle 
\langle \phi_{E} |d_\beta| \phi^{\textrm{d},L}_{i}
\rangle^{*} 
 \right.
 \\\nonumber
 & \;\;\;\;\; \left.
+ 
\langle \phi_{E} |d_\alpha| \phi^{\textrm{d},L}_{i}\rangle 
\langle \phi_{E} |d_\beta| \phi^{\textrm{d},R}_{i} \rangle^* \right]    
\end{align}
Thus, in practice, any expression in our code for the cross section (originally written for variational wavefunctions~\cite{Toffoli:2024:Tiresia})
that contains Hermitian products of two dipole matrix elements 
is transformed, in the non-variational CC case, 
as
\begin{equation} 
%d^{FO}_\alpha  (d^{FO}_\beta)^* \rightarrow \frac{1}{2} 
%\left\{
%d^{FO}_{R,\alpha}
%d^{FO*}_{L,\beta}  + %d^{FO}_{L,\alpha}  %d^{FO*}_{R,\beta} 
%\\
%\revP{
D^{i0}_\alpha  (D^{i0}_\beta)^* \rightarrow \frac{1}{2} \left\{
D^{i0}_{R,\alpha} D^{0i*}_{L,\beta}  + 
D^{0i}_{L,\alpha}  D^{i0*}_{R,\beta}  
\right\} 
%}
\end{equation}
% so that
where $\alpha$ and $\beta$ refer to different Cartesian components $x$, $y$ or $z$.
%which is the form employed in Eq \eqref{eq-A_L}.
%%% SONIA SLUT
We refer to our previous studies~\cite{Moitra:2020,Moitra:2021,Moitra:2022}
for further details.
%~\cite{Oana:2007,Oana:2009,Moitra:2020,Moitra:2021,Moitra:2022}

A step further is to include multielectron effects in the continuum states. A full treatment requires in principle the use of a 
close-coupling approach (equivalent to CI in the continuum), which is expensive and typically restricted to relatively small systems. A practical approximation is to include the response effects induced by the electromagnetic field on the electron cloud. This effect is well described by TDDFT (or RPA in ab initio) and in practice amounts to substituting the bare dipole operator $d_\alpha$ with 
the self-consistent potential (or dressed dipole) operator $V^{\textrm{SCF}}_\alpha$ in the evaluation of dipole matrix elements,~\cite{Zangwill:1980:TDDFT,Stener:2005:B-spline-TDDFT-noniter,
Moitra:2021}
\begin{equation}
    \langle \phi_E | V^{\textrm{SCF}}_{\alpha} | \phi_i\rangle~.
\end{equation}
Despite being computationally much more demanding, continuum TDDFT is still practical for rather large and complex molecules.~\cite{Moitra:2021}

Summarizing, we have three choices for the initial orbital $\phi_i$ in Eq.~\eqref{DipoleMatrixElement}, i.e., DFT and HF {molecular orbital \revS{(MO)}} or EOM-CCSD Dyson, and two for the continuum states, either static DFT or TDDFT. 
The combination of the EOM-CCSD Dyson with the TDDFT continuum provides the most accurate description.
For further computation details, see ESI, 
Section~S2.
% Section~\ref{SI-sec:ComputationalDetails}.

The epichlorohydrin molecule has conformational flexibility, with three stable conformers, denominated g-I, g-II and cis,~\cite{Tam:2006:OR:epichlorohydrin,Stranges:2011} of which g-II is the dominant one.
The three conformers are shown in 
Fig.~S2, ESI.
%Fig.~\ref{SI-%fig:rotamers}, ESI.

All results reported are averages over those obtained for the separate conformers, with appropriate statistical weights (see %Section~\ref{SI-sec:ComputationalDetails}
Section~S1
in ESI for more details).

\section{Results and discussion}

\revS{We remind the reader that epichlorohydrin is a closed shell molecule with 48 electrons,
hence the highest occupied molecular orbital (HOMO) is number 24, HOMO-1 is number 23, and so on.
In the following, we will label the final ionic states by the orbital from which we ionize according to the single determinant description, which is exact for the DFT or HF treatment. 
In the Dyson case, however, the same label will be associated to the actual wavefunction, starting from the lowest-energy ionic state. Thus, the ionisation energies will be labelled as 1A, 2A, 3A, $\cdots$, and the corresponding Dyson orbitals 
(and channels) as 24a, 23a, 22a, $\cdots$, with ``A''/``a'' indicating the symmetry irrep of the ionic state/ionised orbital.}
%
%Recurrence
% \revS{   
% In the following, we will label the final ionic states by the orbital in the single determinant description, which is exact for the DFT or HF treatment, while in the Dyson case the label refers to the actual wavefunction, starting from the lowest-energy ionic state.}

\subsection{Electronic Structure}

Ionization energies and pole strengths computed at the EOM-IP-CCSD/aug-cc-pVDZ level for the three conformers are reported in Table~\ref{tab:IE:Ri:CCSD}, 
\revS{together with
experimental values for the IE's from Ref.~\citenum{Stranges:2011}.
We note that the pole strength values are large and close to 0.9 for all ionizations. 
These should not confused with the height of the peaks in the experimental PE spectra, that depends on the photon energy at which the PE spectra are recorded.}

\begin{table}[H]
\centering
\caption{C$_3$H$_5$OCl. Ionization energies (IE, eV) and intensities (pole strengths $R_i$) computed using EOM-IP-CCSD/aug-cc-pVDZ.
Experimental values from Ref.~\citenum{Stranges:2011}.
\revS{We also indicate, in parentheses, the ``orbital label'' assigned to the Dyson orbital corresponding to each ionisation}.}
\label{tab:IE:Ri:CCSD}
   \begin{tabular}{l|cc|cc|cc|c}
   \hline
 Ionis. $i$ & \multicolumn{2}{c|}{cis} & \multicolumn{2}{c|}{g-I} & \multicolumn{2}{c|}{g-II} &  \\
     \revS{ (Dyson)} %\revS{(orbital)}
    &
    IE & $R_i$ &
    IE & $R_i$ &
    IE & $R_i$ &
    Expt. IE\cite{Stranges:2011}\\
       \hline
1A  \revS{(24a)} 
&    10.29  &  0.8917  &    10.48  &  0.8917  &    10.51  &  0.8906  &  10.56 \\  
2A  \revS{(23a)} 
&    11.00  &  0.9082  &    11.03  &  0.9062  &    11.16  &  0.9096  &  11.28 \\  
3A  \revS{(22a)}
&    11.32  &  0.9052  &    11.13  &  0.9056  &    11.41  &  0.9051  &  11.55 \\  
4A  \revS{(21a)}
&    11.68  &  0.9005  &    11.72  &  0.9013  &    11.65  &  0.9005  &  11.81 \\  
5A \revS{(20a)} 
&    13.49  &  0.9046  &    13.79  &  0.8983  &    13.61  &  0.8970  &  13.54 \\  
6A  \revS{(19a)}
&    13.80  &  0.8922  &    14.03  &  0.8980  &    13.99  &  0.9045  &  14.16 \\  
7A  \revS{(18a)}
&    14.62  &  0.8999  &    14.55  &  0.8977  &    15.13  &  0.8946  &  15.01 \\  
8A  \revS{(17a)}
&    16.06  &  0.8935  &    15.79  &  0.8952  &    15.49  &  0.8941  &  15.64 \\  
9A  \revS{(16a)}
&    16.43  &  0.8876  &    16.67  &  0.8861  &    16.59  &  0.8861  &  16.49 \\  
10A  \revS{(15a)}
&   17.68  &  0.8791   &   17.68  &  0.8805  &    17.70  &  0.8821  &  17.47 \\  
\hline
\end{tabular}
\end{table}
As can be seen, the change from one conformer to the other is not negligible, up to a few tenths of an eV. The ionisation energies computed for the dominant g-II conformer compare pretty well with the experimental ones. HF and DFT eigenvalues are reported in 
Table~S1 (ESI).
%Table~\ref{SI-tab:hf-dft-mo} (ESI).
\revS{The Dyson orbital composition} in terms of HF MOs is displayed in Table~\ref{tab:weights:CCSD}. 
%Table~\ref{SI-%tab:Mulliken} 
Table~S2 
reports the atomic 2p (and Cl 3p) populations of the HF MO (other populations, with the exception of H 1s, are very small and therefore omitted). 

\begin{table}[htb!]
\centering
\caption{Epichlorohydrin. Highest coefficients $\gamma_p^L %\times 
\gamma_p^R$ for each ionization channel. 
Numbers in parentheses labeled `HF' denote the Hartree-Fock molecular orbitals contributing to the given ionization/Dyson orbital.
Only contributions greater than 0.1 are reported.}
\label{tab:weights:CCSD}
\begin{tabular}{l|c|c|c}
    \hline
Ionis. & cis & g-I & g-II \\
(Dyson) &  &   &   \\\hline
\hline
1A (24a) & 
\ 0.198(HF 21a) &
0.304(HF 21a) & 
0.376(HF 21a)\\
    & 
+0.108(HF 22a) &
+0.250(HF 22a) &
+0.455(HF 23a) 
\\
         & 
+0.267(HF 23a) &
+0.244(HF 24a) & 
\\
         & 
+0.307(HF 24a)
         &&\\
%REORGANISING
%
% 1A (24a) & 
% 0.198(HF 21a) 
% %& & & \\
% %&& 
% +0.108(HF 22a)+ & 0.304(HF 21a)+0.250(HF 22a)+ & 0.376(HF 21a)+0.455(HF 23a)\\
% & 0.267(HF 23)+0.307(HF 24a) & 0.244(HF 24a) & \\
\hline
2A (23a) & 
0.373(HF 23a) 
&
0.312(HF 23a)
&
0.884(HF 24a) \\
&
+0.513(HF 24a) & 
+0.526(HF 24a) & \\\hline
%%%
% 2A (23a) & 0.373(HF 23)+0.513(HF 24) & 0.312(HF 23a)+0.526(HF 24a) & 0.884(HF 24a)\\\hline
%
3A (22a) 
& 0.133(HF 21a)
& 0.110(HF 21a)
& 0.425(HF 21a)
\\
&
+0.461(HF 22a)&
+0.151(HF 22a)&
+0.424(HF 23a)
\\
&+0.238(HF 23a) 
&+0.510(HF 23a)
&
\\
&
+0.132(HF 24a)
& 
\\
\hline
%%%%%%%%%
% 3A (22a) & 
%  0.133(HF 21a)
% +0.461(HF 22a)& +0.110(HF 21a)
% +0.151(HF 22a)& 
% +0.425(HF 21a)
% +0.424(HF 23a)
% \\
% &+0.238(HF 23a) 
% & 0.510(HF 23a)
%  +0.132(HF 24a)
% & %0.425(21)+0.424(23)
% \\
%\hline
4A (21a) 
& 0.544(HF 21a)
%+0.313(HF 22a) 
& 0.438(HF 21a)
%+0.455(HF 22a) 
& 0.822(HF 22a)
\\
 & 
+0.313(HF 22a) &  +0.455(HF 22a) & 
\\
\hline
5A (20a) 
& 0.883(HF 20a) 
& 0.834(HF 20a) 
& 0.874(HF 20a)
\\
\hline
6A (19a) 
& 0.885(HF 19a) 
& 0.114(HF 18a)
% +0.718(HF 19a) 
& 0.887(HF 19a)
\\

&   
&  
 +0.718(HF 19a) 
& 
\\\hline
7A (18a) 
& 0.871(HF 18a) 
& 0.773(HF 18a)
%+0.108(19) 
& 0.838(HF 18a)
\\
&  
& +0.108(HF 19a) 
& 
\\\hline
8A (17a) 
& 0.861(HF 17a) 
& 0.885(HF 17a) 
& 0.850(HF 17a)
\\\hline
9A (16a) 
& 0.844(HF 16a)
& 0.874(HF 16a) 
& 0.872(HF 16a)\\
\hline
\end{tabular}
\end{table}

%As can be seen, 
The Dyson orbitals 
\revS{of the first four ionised states}
show a heavy mixing of the four outermost 
\revS{(e.g., 24a, 23a, 22a and 21a)} Hartree-Fock MO's, while those of the inner ones 
%are not, 
{have mostly pure, unmixed character},
except minor mixing for \revS{6A (19a) and 7A (18a)} in g-I. Also, 
HF MO's populations change significantly from one conformer to the other, although Cl 3p always dominates orbitals 24a and 23a, which are essentially Cl lone pairs.
They seem to indicate a corresponding softness of the orbital composition, such that even modest conformational changes induce significant variations in individual orbitals. 
The final Dyson orbitals,
on the other hand,
appear more robust, as evidenced by the relatively modest changes in the $\beta$ profiles evaluated for the three conformers in 
Fig.~S3.
%Fig.~\ref{SI-fig:beta_Rotamers_TDDFT}.

\subsection{Photoelectron angular distribution measurements}

\begin{figure}[h]
 \centering
\includegraphics[width=0.95\linewidth]
{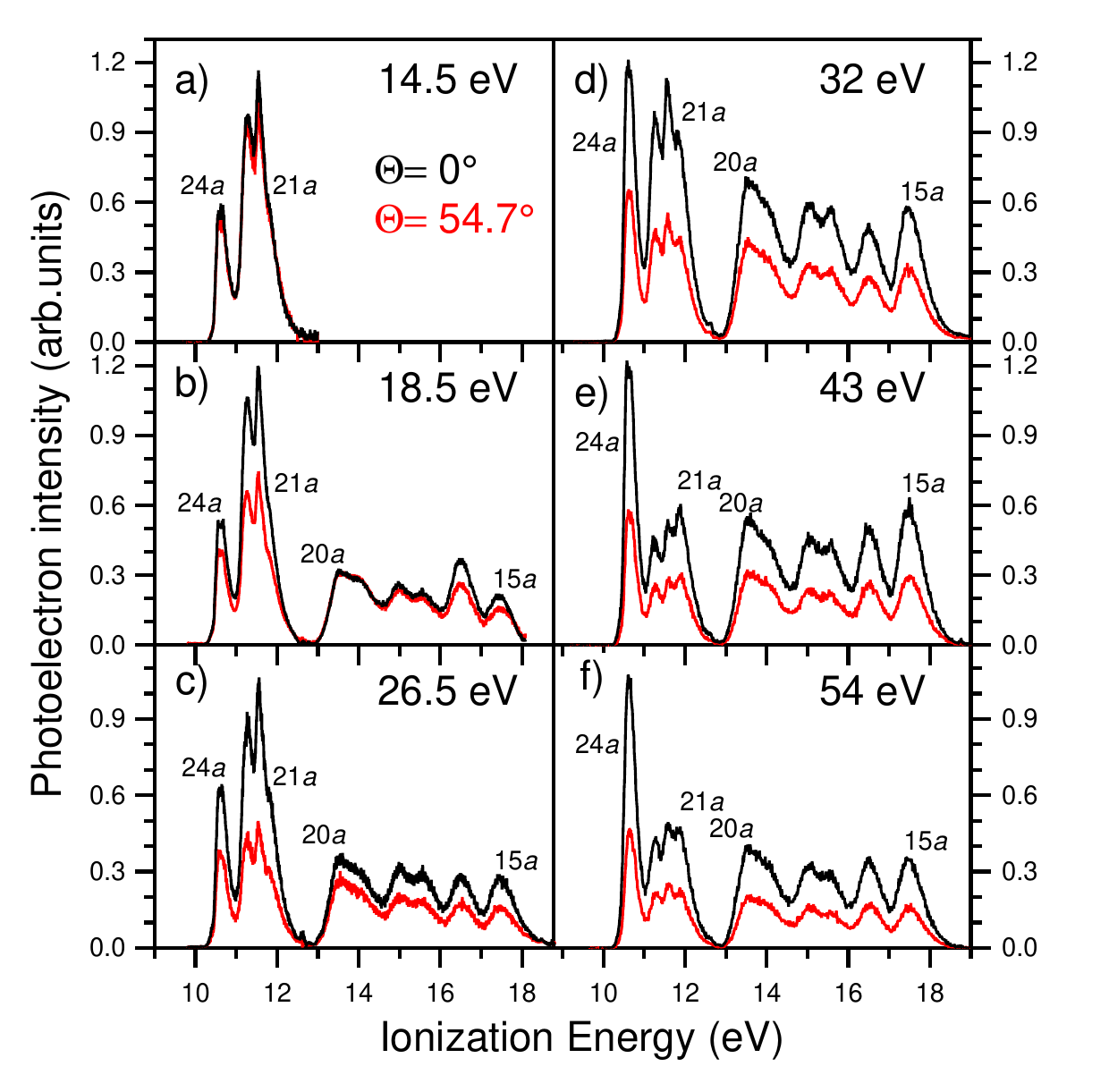}
   \caption{Experimental outer valence PE spectra of epichlorohydrin recorded at the magic angle $\theta = 54.7^\circ$(red) and $\theta = 0^\circ$ (black), and at different photon energies (panels a-f).}
   \label{fig:experiment}
 \end{figure}

The anisotropy parameters $\beta_i(h\nu)$ associated with epichlorohydrin's outer valence PhotoElectron (PE) bands observed in the 10--19 eV IE range, have been studied experimentally and theoretically as a function of photon energy. Pairs of PE spectra recorded at two different angles, $\theta = 0^\circ$ and $\theta = 54.7^\circ$, from which the $\beta$ values have been derived {(see Section~\ref{Methods} 
and 
Section~S1, ESI,
%Section~\ref{SI-sec:extraction}, ESI,
for details)}, 
are reported in Fig.~\ref{fig:experiment} for a representative set of photon energies. 

\subsection{Anisotropy $\beta$}

\begin{figure}[hbpt!]
    \centering
\includegraphics[width=0.95\linewidth]{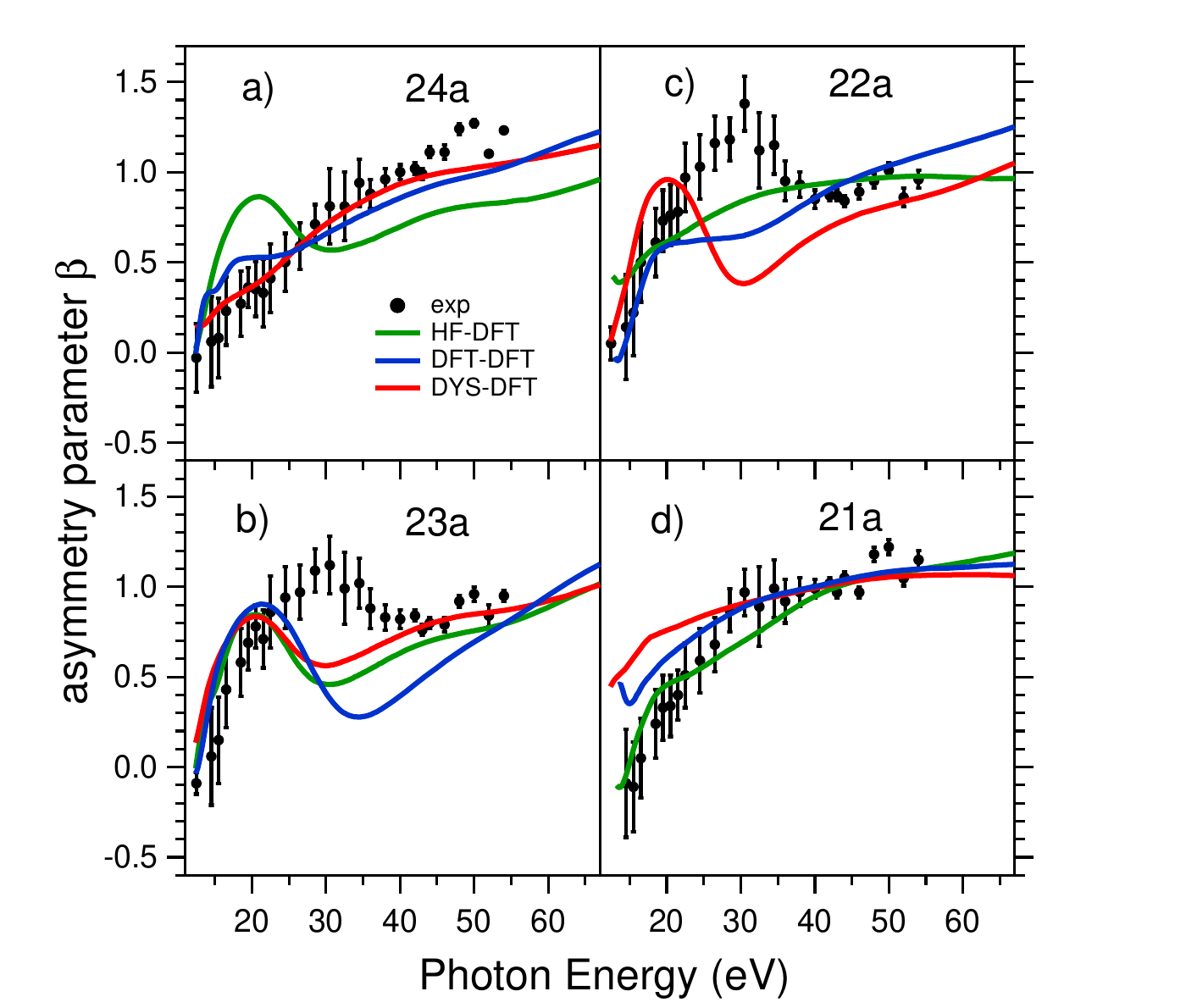}
    \caption{Asymmetry parameters $\beta_i$ as a function of the photon energy for orbitals \revS{(channels)} 24a, 23a, 22a and 21a. Experiment versus computational results obtained using a DFT representation of the continuum together with Hartree-Fock (label HF-DFT), Density Function Theory (label DFT-DFT) and EOM-CCSD Dyson (label DYS-DFT) orbitals.}
    \label{fig:BetaDFT}
\end{figure}

\begin{figure}[hbpt!]
\centering
\includegraphics[width=0.95\linewidth]{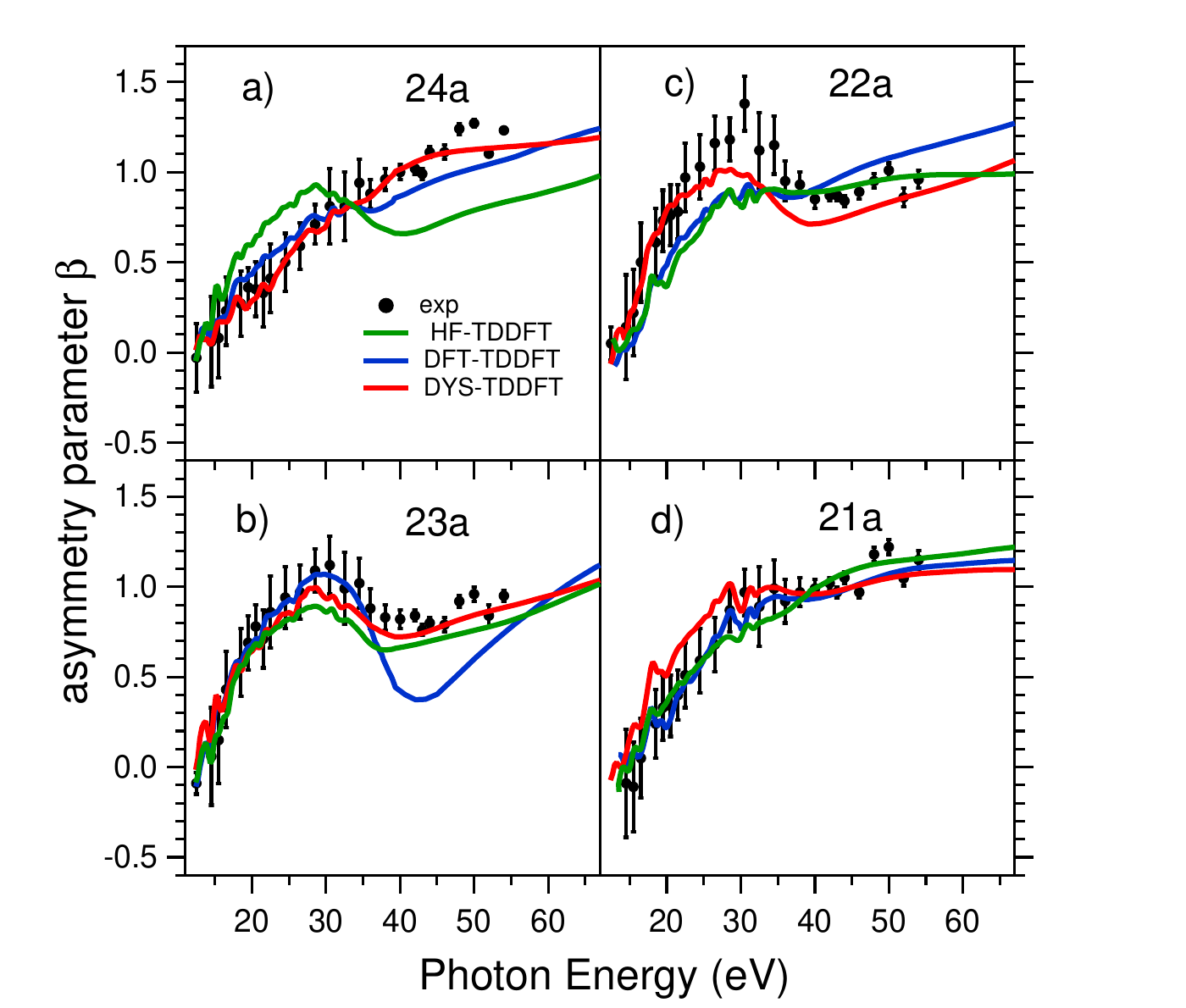}
\caption{Asymmetry parameters $\beta_i$ as a function of the photon energy for orbitals \revS{(channels)} 24a, 23a, 22a and 21a. Experiment versus computational results obtained using a TDDFT representation of the continuum together with Hartree-Fock (HF-TDDFT), Density Function Theory (DFT-TDDFT) and EOM-CCSD Dyson (DYS-TDDFT) orbitals.}
\label{fig:BetaTDDFT}
\end{figure}

The experimental $\beta_i(h\nu)$ values of the outermost valence ionizations 24a--21a \revS{} are reported in Fig.~\ref{fig:BetaDFT} and Fig.~\ref{fig:BetaTDDFT} (black dots in panels a-d),
{together with the simulated results. The latter differ in the continuum description, being DFT in Fig.~\ref{fig:BetaDFT} and TDDFT in Fig.~\ref{fig:BetaTDDFT}}. 
% \revS{Note that we label final ionic states by the orbital in the single determinant description, which is exact for the DFT or HF treatment, while in the Dyson case the label refers to the actual wavefunction, starting from the lowest-energy ionic state.}

The dynamical behaviors of the experimental $\beta$ parameters, shown in Fig.~\ref{fig:BetaDFT}, display clear similarities in the low energy region. 
%
% \begin{figure}
% \centering
% \includegraphics[height=10cm]{Figures/Fig_a_Beta_DFT_continuum_layout.pdf}
% \caption{ }
% \end{figure}
%
All the $\beta$ values are slightly negative or nearly zero at threshold, and rapidly increase with photon energy to reach, at 30 eV, largely positive values of approximately $+1.0$. 
A steeper increase is observed for the 23a and 22a channels with a change in slope at around 32 eV. Here, a maximum is located which corresponds to $\beta$ values of +1.1 (23a) and +1.35 (22a). These two curves are the only ones exhibiting an oscillatory behaviour. This is characterized by a high-energy minimum which is measured for both channels at approximately 43 eV (see panels b and c in Fig.~\ref{fig:BetaDFT} and Fig.~\ref{fig:BetaTDDFT}). 
Worth noting is the difference between the two channels in the $\beta$ oscillation amplitude, namely the maximum-to-minimum $\beta$ change, these amplitudes being 0.3 and 0.5 for 23a and 22a, respectively. A change in slope in the $\beta$ curve at 32eV is also observed for the other two photoionization processes 24a and 21a. 
While the slope change is modest for the 24a channel, the change observed is instead large in the 21a curve, where a clear plateau appears in the 30--43 eV photon energy region with $\beta$ = +1.0. The 
$\beta$ increase at low energies for the first four PE bands is also evident in 
Fig.~\ref{fig:experiment}. 
Here, isotropic $\beta$ values, $\beta$=0, are measured at 14.5 eV (panel a), as the spectra recorded at 0$^\circ$ and 54.7$^\circ$ almost perfectly coincide. In going from 14.5 to 32eV, from panel a to panel d, the relative intensities of these bands in the 0$^\circ$ spectra clearly display a gradual increase, which corresponds to the $\beta$ monotonic increase quoted in the panels of 
Fig.~\ref{fig:BetaDFT} at low energies.
The $\beta_i(h\nu)$ parameters have also been measured for the other outer valence ionizations 
20a--15a (see ESI). They do not exhibit an oscillatory behaviour and show an increase similar to those of 24a and 21a channels, reaching values in the range 0.9-1.2 at 54 eV photon energy.

In Fig.~\ref{fig:BetaDFT} are also shown the computational results relative to the same DFT continuum, but with different initial orbitals, DFT, HF and EOM-CCSD Dyson (labelled DYS). The marked differences show the sensitivity of the calculated asymmetry parameter to the bound-state description, and in particular between HF and DFT orbitals, which are generally believed to be quite close to each other. 
The most notable difference  is 
\revX{the position of the maximum and of the oscillation expected from the CM effect}. With HF orbitals, the \revX{CM feature} appears strong in the HOMO (24a) and in the 23a channel, with comparable amplitude. DFT shows instead a small hint of it in 24a, another inflection in 22a, and a very strong oscillation in 23a. 
The use of Dyson orbitals correctly displays \revX{the CM feature} in the 22a and 23a channels, and follows the experimental trend with a larger amplitude in 22a compared with 23a. In all cases 
\revX{the energy positions of the maximum and the minimum} are strongly shifted towards threshold, by almost 10 eV. For 21a all approaches are qualitatively correct, \revX{with no sign of CM feature}. Here, HF gives the best agreement at low energy, while both DFT and Dyson results are overestimated; then at higher energy the shape favours the latter. A good agreement with experiment is obtained from the Dyson results for the HOMO (24a), and for the steep rise at low energy in channels 22a and 23a. Due to the energy shift of the CM feature, it is difficult to argue for the intermediate energy region. 

% \begin{figure}[hbpt!]
%   \centering
%   \includegraphics[width=0.7\linewidth]{}
%   \caption{Asymmetry parameters $\beta_i$ as a function of the photon energy for orbitals 24, 23, 22 and 21. Experiment versus computational results obtained using a TDDFT representation of the continuum together with Hartree-Fock (HF-TDDFT), Density Function Theory (DFT-TDDFT) and Coupled cluster Dyson (DYS-TDDFT) orbitals.}
%   \label{}
% \end{figure}

TDDFT continuum strongly
improves \revX{the position of the CM feature}, as is displayed in Fig.~\ref{fig:BetaTDDFT} for the same three descriptions of the bound states.
The improvements brought about by the TDDFT continuum are, in fact, dramatic.  This has been observed long ago in the case of rare gas atoms,~\cite{Zangwill:1980:TDDFT,Stener:1995:TDDFT} and in small molecules containing second row elements, like PH$_3$ or Cl$_2$.~\cite{Stener:2000:TDDFT,Stener:2005:B-spline-TDDFT-noniter}
In the present case, the agreement of the EOM-CCSD Dyson/TDDFT approach is almost quantitative. The \revS{energy positions} of the CM maximum and minimum \revS{are} now correct, as 
\revS{correct} is the range of the oscillation, perhaps only slightly underestimated for the 22a ionization. Also the overall shapes, including the shape of the low energy rising leg and the high energy behaviour, are very well reproduced. Even the behaviour of the other curves (DFT and HF) is improved, due to the correct energy position of the CM feature, although the shortcomings already noted, i.e., the misplacing of the CM feature in 24a by the HF orbitals, the overestimate of it in 23a by DFT, and the underestimate by both HF and DFT in 22a, are of course unchanged. 
The high energy slope is also not well reproduced in some cases.

To summarize, the use of the TDDFT continuum 
is paramount to obtain the correct location \revX{of the CM feature}, and it generally improves the overall description of the $\beta$ profiles. 
\revX{The} location and amplitude of the oscillation are entirely due to the nature of the initial orbitals, and specifically their AO composition, and give a very clear experimental signature of the orbital rotation effect upon ionization theoretically predicted.~\cite{VonNiessen:1982:hole-mixing}

\subsection{Partial cross sections}
\begin{figure}
\centering
\includegraphics[width=0.95\linewidth]{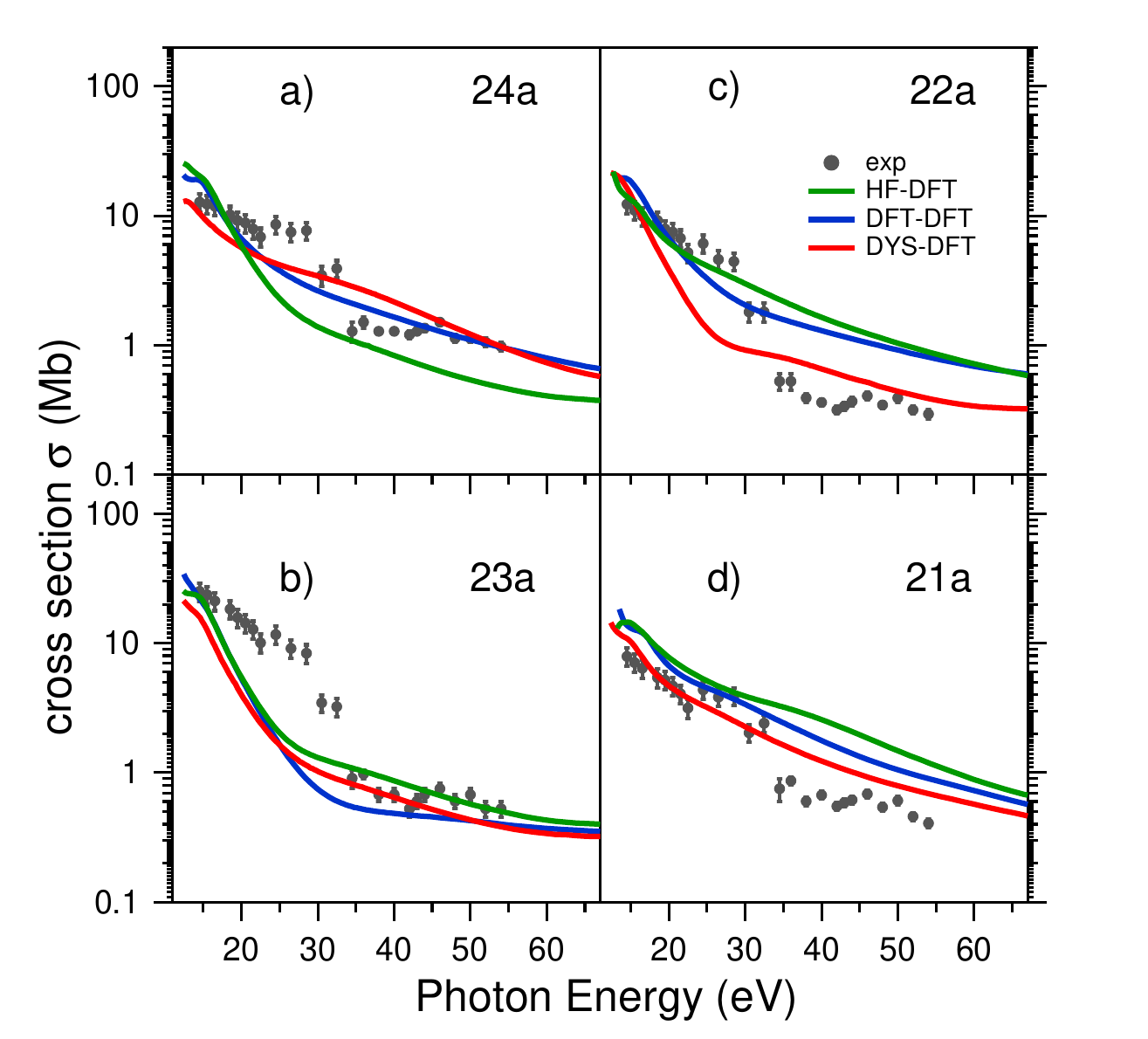}
\caption{Cross section $\sigma$ (Mb, logarithmic scale) as a function of the photon energy for orbitals 24, 23, 22 and 21. Experiment versus computational results obtained using a DFT representation of the continuum together with Hartree-Fock (HF-DFT), Density Function Theory (DFT-DFT) and EOM-CCSD Dyson (DYS-DFT) orbitals.}
\label{fig:sigmaDFT}
\end{figure}

The behaviour of the cross sections tells a similar story, although less vivid. As before, experimental results and theoretical curves relative to the DFT continuum are reported in Fig.~\ref{fig:sigmaDFT}.
%%%

\begin{figure}
\centering
\includegraphics[width=0.95\linewidth]{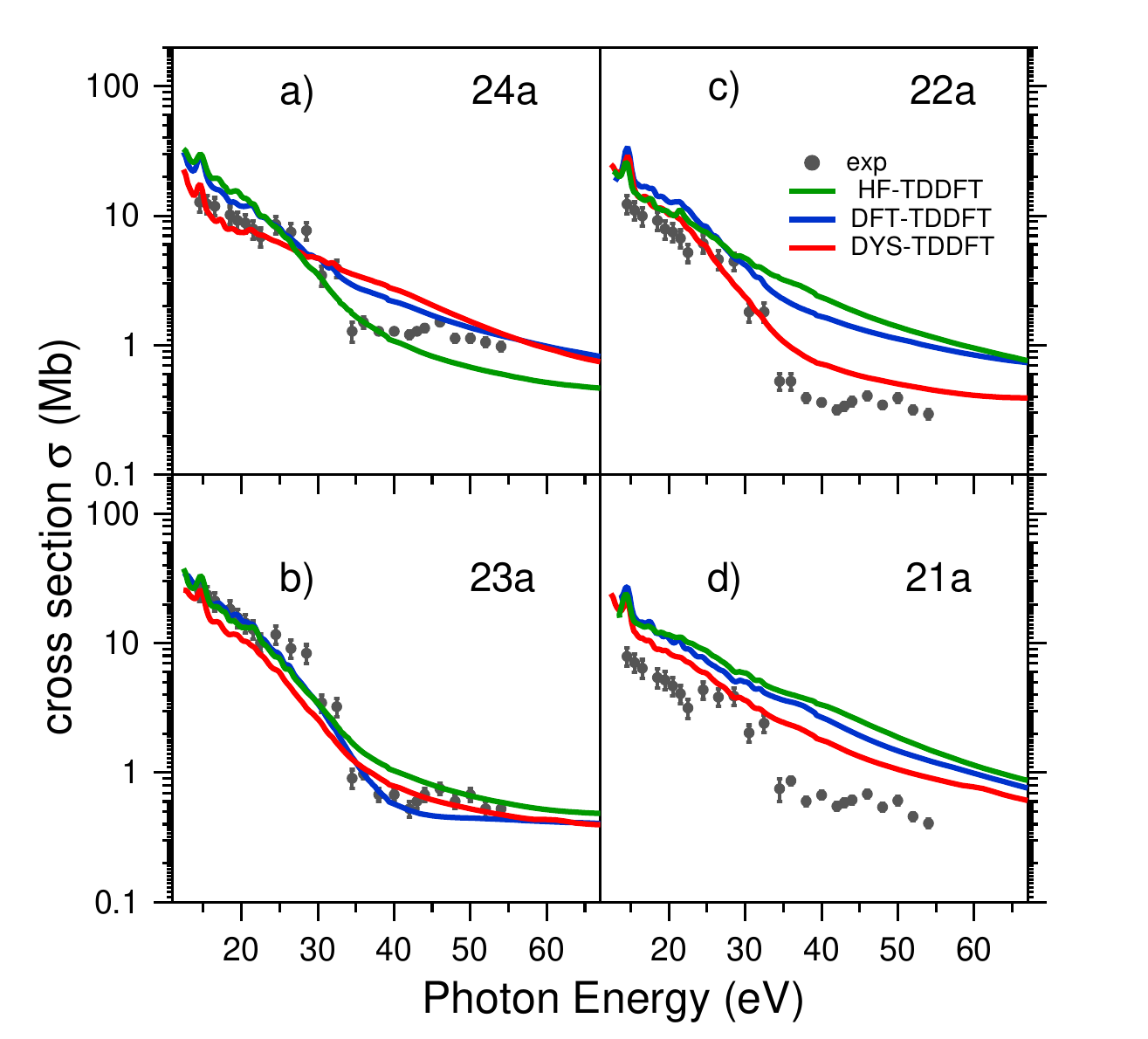}
\caption{Cross section $\sigma$ (Mb) as a function of the photon energy for orbitals 24a, 23a, 22a and 21a. Experiment versus computational results obtained using a TDDFT representation of the continuum together with Hartree-Fock (HF-TDDFT), Density Function Theory (DFT-TDDFT) and EOM-CCSD Dyson (DYS-TDDFT) orbitals.}
\label{fig:sigmaTDDFT}
\end{figure}

%%%
Again, two different behaviours are observed experimentally for channels 22a and 23a, where a sudden drop in $\sigma$ is seen in the region of the CM, around 40 eV, and for 21a, 24a, where the decrease is smoother. Note also the slight (log scale) but well visible recovery of the cross section after the CM. As before, HF results show the step-like decline in the HOMO 24a while it misses it in 22a. DFT also misses it in 22a while it exaggerates it in 23a, basically shifting most Cl 3p contribution to this ionization. 
The cross-section decrease in 21a is smoother in all approaches, and the same occurs for DFT and Dyson also for 24a. One could expect that the CM feature were propagated from 23a, 22a, where it is native due to the Cl 3p participation, to the neighboring channels due to interchannel coupling, absent in DFT but included in the TDDFT continuum. 
However, this is not borne out, as seen in Fig.~\ref{fig:sigmaTDDFT}, where TDDFT vastly improves the description of 22a and 23a while the effect is less marked for 24a and 21a, as TDDFT mostly affects the CM feature.
%%%

%%%
There is no sign of the cross section intensity recovery after the minimum. Since this feature is more or less similar in all experimental channels, it may not be significant, and no clear signs of interchannel coupling effects appear also in the experiment.
Clearly, the overall final agreement is less good than for the $\beta$ values. Moreover, experiment seems to indicate more structure than is theoretically predicted. 
It should be mentioned that the experimental measurement of cross sections is intrinsically more difficult than that of the $\beta$ parameters, and that the CM effect in cross sections manifests itself as a much weaker relative intensity change. As a consequence, it is more difficult to be described theoretically.
Since only relative cross sections have been experimentally measured, a conversion to absolute scale was required to make a direct comparison between experiment and theory possible (see Fig.~\ref{fig:sigmaDFT} and Fig.~\ref{fig:sigmaTDDFT} 
).
The conversion factor was determined by comparing the sum of the experimental cross sections of the 24a ionization channel 
measured at all the selected photon energies to the sum of the corresponding theoretical values calculated by the Dyson-TDDFT method.

% In other words, the experimental values were normalized in such a way that the sum of the experimental values measured at the different photon energies for channel 24a was identical to the sum of the Dyson-TDDFT theoretical values calculated for the same photon energies. All the other channels were then scaled by the same normalization factor.

% \revS{[TO STEFANO: could you please recheck the bit above about the conversion factor? It is not 
% so clear what you did....]}
% One has to notice nonetheless that extraction of experimental cross section is significantly more difficult and some minor 
% %structures 
% features
% can be spurious. 
%Moreover, only relative cross sections can be extracted so that experimental results have been normalized  as to obtain the best agreement with the DYS-TDDFT results for the 23a channel \revS{RIGHT, STEFANO?}.

\begin{figure}
    \centering
    \includegraphics[width=0.95\linewidth]{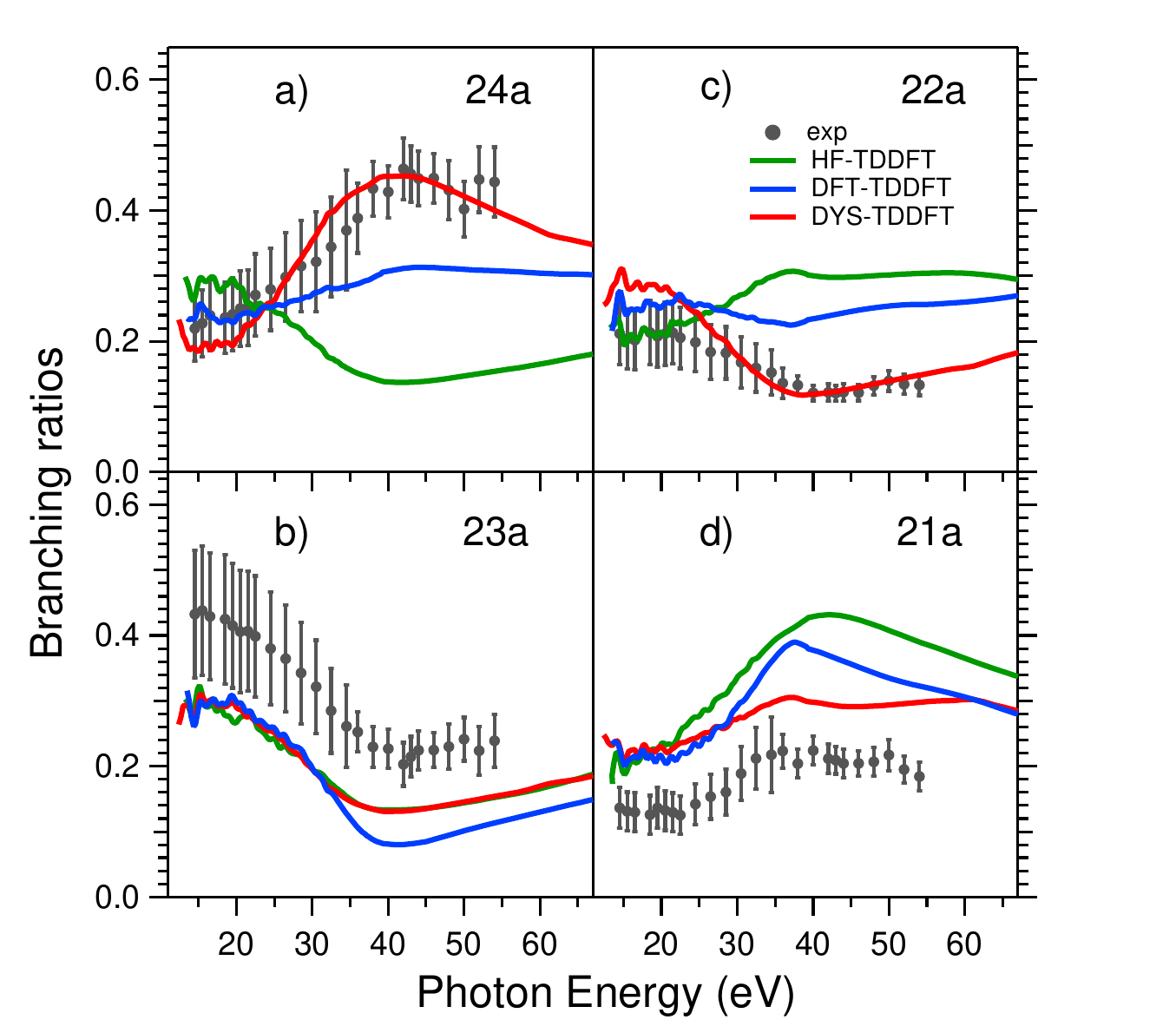}
    \caption{Branching ratios for orbitals 24a, 23a, 22a and 21a as a function of the photon energy. Experiment versus computational results obtained using a TDDFT representation of the continuum together with Hartree-Fock (HF-TDDFT), Density Function Theory (DFT-TDDFT) and EOM-CCSD Dyson (DYS-TDDFT) orbitals.}
    \label{fig:branchingTDDFT}
\end{figure}

An alternative
presentation of the comparison between experimental and theoretical cross sections,
which eliminates the ambiguity in the absolute values, is the use of branching ratios (BR), i.e. in the present case the ratio of individual channel cross section to their sum relative to the four channels considered.  Experimental and theoretical values relative to the TDDFT continuum are reported in Fig.~\ref{fig:branchingTDDFT}, and those corresponding to the DFT continuum are available in the ESI (see Fig.~S7).
%\ref{SI-fig:BR:24-21:DFT}).
%
% \begin{figure}[hbpt!]
% \centering
% \includegraphics[width=0.95\linewidth]{Figures/Fig_e_BR_24a_21a_exp_TDDFT_Continuum.pdf}
% \caption{Branching ratios for orbitals 24, 23, 22 and 21 as a function of the photon energy. Experiment versus computational results obtained using a TDDFT representation of the continuum together with Hartree-Fock (HF-TDDFT), Density Function Theory (DFT-TDDFT) and Coupled Cluster Dyson (DYS-TDDFT) orbitals.\label{fig:branchingTDDFT}}
% \end{figure}

The drawback of the BR is that a strong resonant feature present in one channel influences all the others, so that even a channel with a smooth cross section will display induced features.  The upside is that spurious fluctuations in the cross sections cancel out, and the profiles are cleaner and independent on normalization.
In Fig.~\ref{fig:branchingTDDFT} the %discrimination 
difference
between the 
alternative descriptions is very clear, and the same features already discussed  for each channel appear prominently. An excellent description is provided by the 
DYS-TDDFT approach for the 24a and 22a channels. For the 23a and 21a ones,
the shape is very well reproduced, but the ratio is underestimated for 23a and correspondingly overestimated for 21a. 
While an obvious suggestion is a difficulty in the deconvolution of partially overlapping bands, of different widths and shape, we do not have presently an unequivocal attribution of this disagreement either to a deficiency of the processing of experimental data or the inadequacy of the theoretical approach.

A brief analysis of the remaining ionizations, 15a-20a is presented in ESI, 
see Section~S3.
%see Section~\ref{SI-sec:otherchannels}.

\section{Conclusions}
Photoionization cross sections $\sigma$ and angular distributions $\beta$ have been accurately measured for the entire outer valence region of the chiral molecule epichlorohydrin over an extended energy range, which is important to fully characterize specific signatures. Prominent features in the $\beta$ parameters appear in two of the four outermost ionizations, attributed to Cooper Minimum effects linked to a Cl 3p participation in the relevant orbitals. Detailed theoretical calculations employing DFT, HF and Dyson orbitals computed at the EOM-CCSD level provide different energy positions of the Cooper Minimum features in the $\beta$ parameters, 
a clear signature of important correlation effects upon ionization, only observable in transition properties like photoionization dynamical parameters. 
Such effects, predicted long ago but never observed, do not show up in \revS{ionisation energies} or pole strengths, which remain close to 1, so they are not due to a \textit{breakdown of the one-particle picture} \revS{(as defined by Cederbaum)}, but represent an \textit{orbital mixing}, or \textit{orbital rotation upon ionization} caused by differential relaxation and correlation upon ionization. 
The use of the TDDFT continuum is essential for the proper 
%location 
energy prediction
\revS{(i.e, its position on the energy axis)} of the 
Cooper Minimum and, in general, \revS{coupling the TDDFT continuum to the Dyson orbital description of the bound initial and final states} provides a very accurate description of the experimental results. Although less prominent, such \revS{orbital mixing} effects also manifest themselves in the cross section and in the branching ratios. While their shapes are very well described by theoretical results, a discrepancy in some absolute values is apparent, whose origin is not clear at present. Additional discrepancies between DFT and HF initial orbitals are also of interest, and clearly highlighted by the present results, which are shown to be highly informative on the electronic structure of the molecule. 
Although less dramatic, such effects are also seen in the remaining part of the valence shell.

\revS{
% SENTENCE MOVED FROM THE INTRODUCTION. FOLD IT INTO THE MANUSCRIPT...

To conclude we note that, 
because of the lack of symmetry and the dense manifold of valence orbitals in chiral molecules, such orbital mixing effects are expected to be relatively common in their photoionization properties, and in particular in their PECD spectra, as already observed in Ref.~\citenum{Waters:2022:cphc}, for which the present methodology is expected to provide an accurate description. 
Although the dichroic parameter $\beta_1$ decreases fast with increasing electron kinetic energy and is therefore a less sensitive probe of the CM, it will probably be strongly affected by the orbital rotation.~\cite{Waters:2022:cphc}
Since theoretical simulations of PECD, like in conventional CD, are almost mandatory to interpret the observed signal, the present study has strong implication for future PECD studies.
Further investigations in this regard are in progress.
}

\section*{Acknowledgments}
The authors gratefully acknowledge the staff of ELETTRA for the smooth operation of the storage ring during the measurements. 
% S.S. thanks Sapienza University for financial support through Progetto di Ateneo 2016 (prot. n. RG116154C8E02882). 
S.S. thanks Sapienza University for financial support through Progetto Ateneo Medio 2023, RM123188F6672F66
L.S. acknowledges financial support by IOM-CNR (research grant), SBAI Department of Sapienza (PhD grant) and Sapienza University for a research grant (prot.n. AR11715C81FF05FF).
S.C. thanks
the Hamburg  Cluster of Excellence ``CUI: Advanced Imaging of
Matter'' for the 2024 Mildred Dresselhaus Prize.
T.M. and S.C. acknowledge the Marie Sk{\l}odowska-Curie European Training Network COmputational
Spectroscopy In Natural sciences and Engineering (COSINE), Grant Agreement No. 765739.

\section*{Supporting Information}
S1 Spectral extraction of the experimental $\beta$ parameters; 
S2 Computational Details; S3 Higher energy ionizations (deeper ionisations); 
S4 Cartesian coordinates of the three rotamers (Bohr units);
S5 Additional Tables and Figures;
S6 Comparison between experimental and theoretical observables;
S7 Derivation of $\beta_\textrm{av}$ over conformers.

\section*{Data Availability}

In addition to the details given in the manuscript and ESI file,
all data underlying this study are available from the authors upon reasonable request.

\section*{References}
\bibliography{rsc,stefano}

\end{document}

\maketitle

\clearpage
\section{Spectral extraction of the experimental $\sigma$ and $\beta$ parameters}
\label{sec:extraction}

To extract the $\sigma(h\nu)$ and $\beta_i(h\nu)$ values from the spectrum, the relative intensities of the single PE components in terms of peak areas were obtained by the global multipeak fitting procedure~\cite{GlobalMultipeak} described below.
First, the PE spectra were normalized by the photon flux and corrected by the proper spectrometer transmission function. 
The background contribution was then systematically removed from all the spectra using the following procedure. The background, as described by an \texttt{arctan} 
function plus a linear function, was determined by 
a $\chi$-square based fitting of spectra recorded without epichlorohydrin, and subtracted from the total spectrum. The initial part of the step onset of this function was involved in the procedure. This subtraction, although applied to all spectra, was actually needed for a subset of PE spectra recorded at low photon energies, and proved especially significant in the low kinetic energy part of the spectrum.
The net spectra, thus obtained, were analysed by a global multipeak fitting method,~\cite{GlobalMultipeak} which simultaneously involved the data obtained at all investigated photon energies, as well as at 
the 0$^\circ$ and 54.7$^\circ$
detection angles.
Ten spectral components were considered, namely the PE bands associated with the ionizations from the 24a--15a MOs. The PE spectrum could be conveniently divided in two not overlapping regions. The first group of four bands, 
the 24a--20a components in the 
10.0--13.0 eV IE range, could be analysed separately from the other bands. These four bands exhibit a variety of shapes and intensities, as well as different extent of overlap, as the photon energy is increased. In order to describe their envelopes and ionization potentials precisely and consistently, a program was developed which efficiently makes curve fits of the spectral data. Because all bands have an asymmetric shape, with longer tail on the high IE side (right side), an asymmetric Gaussian was adopted, namely a set of bi-Gaussian functions were used for the spectral deconvolution. 
Such an asymmetric function is made of two half Gaussian functions which share the summit of the peak. Four parameters fully define the function, the location of the summit (E$_0$, the vertical IE value), the standard deviations of the left and right half Gaussian functions $\sigma_l$ and $\sigma_r$, and the scale factor $\delta$ this latter being related to the peak area ($\delta(\sigma_l+\sigma_r)/2$) and the peak height 
($\delta/\sqrt{2\pi}$). This method~\cite{Yu:2010,Buys:1972} 
%[2, 3] 
has been successfully applied to analyze experimental PE spectra.~\cite{Boccia:2008,Lichtenberger:1990,Lichtenberger:1976} 
The four fit parameters defining the PE band positions, namely the vertical IEs, were considered differently. The IE of the first band (24a MO) was set as free parameter in each PE spectrum, whilst the positions of the other three bands were defined as relative IEs referred to the first HOMO PE peak and considered equal in all the spectra, since the relative positions of the PE bands are actually constant in all the spectra. 
The eight left and right standard deviation free parameters, which define the peak shape of the four PE bands, were considered constant in all the spectra, namely each PE peak preserves its own shape ($\sigma_l$  and $\sigma_r$ values) on changing the photon energy and the angle of detection. The scale factors $\delta_i$ of the four PE bands were set as free fit parameters and specific for each PE spectrum. The global multipeak fit method~\cite{GlobalMultipeak} thus allowed the PE band relative intensities to be measured in terms of peak areas, and provided a systematic method to measure the $\beta_i(h\nu)$ parameters and the relative photoionization cross sections $\sigma_i(h\nu)$, 
for the valence photoionization processes of epichlorohydrin as a function of photon energy.
The six PE bands at higher energies, the 
20a--15a spectral components, were analysed by a simpler procedure, since all these bands share, with good approximation, the same peak shape. The whole set of the PE experimental spectra, namely recorded at all the investigated photon energies and the 
0$^\circ$ and 54.7$^\circ$ detection angles, were simultaneously involved in the global multipeak fit analysis. 
Similarly to the first group of bands, the six peak positions were referred to the vertical IE of the HOMO band and considered as six free fit parameters, whose values are preserved in all the spectra. Here, only two standard deviation free parameters, 
 $\sigma_l$ and $\sigma_r$, were sufficient to accurately describe the peak shape of the six bands. This peak shape was considered constant, being not significantly affected by the change in photon energy and angle of detection. The six scale factors, $\delta_i$, were defined as free fit parameters and specific for each PE spectrum. The best fit scale factors $\delta_i$ and the standard deviations $\sigma_l$ and $\sigma_r$, obtained by the global fit spectral analysis, allowed us to measure the $\beta_i(h\nu)$ and $\sigma_i(h\nu)$ photoionization observables as a function of photon energy also for the high energy PE bands of epichlorohydrin.
The high flexibility of the fitting package of IGOR PRO~\cite{GlobalMultipeak} allowed us to provide the bi-Gaussian function as external (user-defined) model fit function to curve fit a large set of multicomponent experimental PE spectra, simultaneously, and to define the nature, constraints, and links to freely chosen experimental data subsets of the fit parameters, according to the description given above. The large change in relative intensity of the first four PE bands, observed in the investigated photon energy range, was fruitfully used, by means of the global multipeak fit analysis,~\cite{GlobalMultipeak} to shed more light in this congested PE spectral region.

\begin{figure}[hbpt!]
    \centering
     \includegraphics[width=0.8\textwidth]{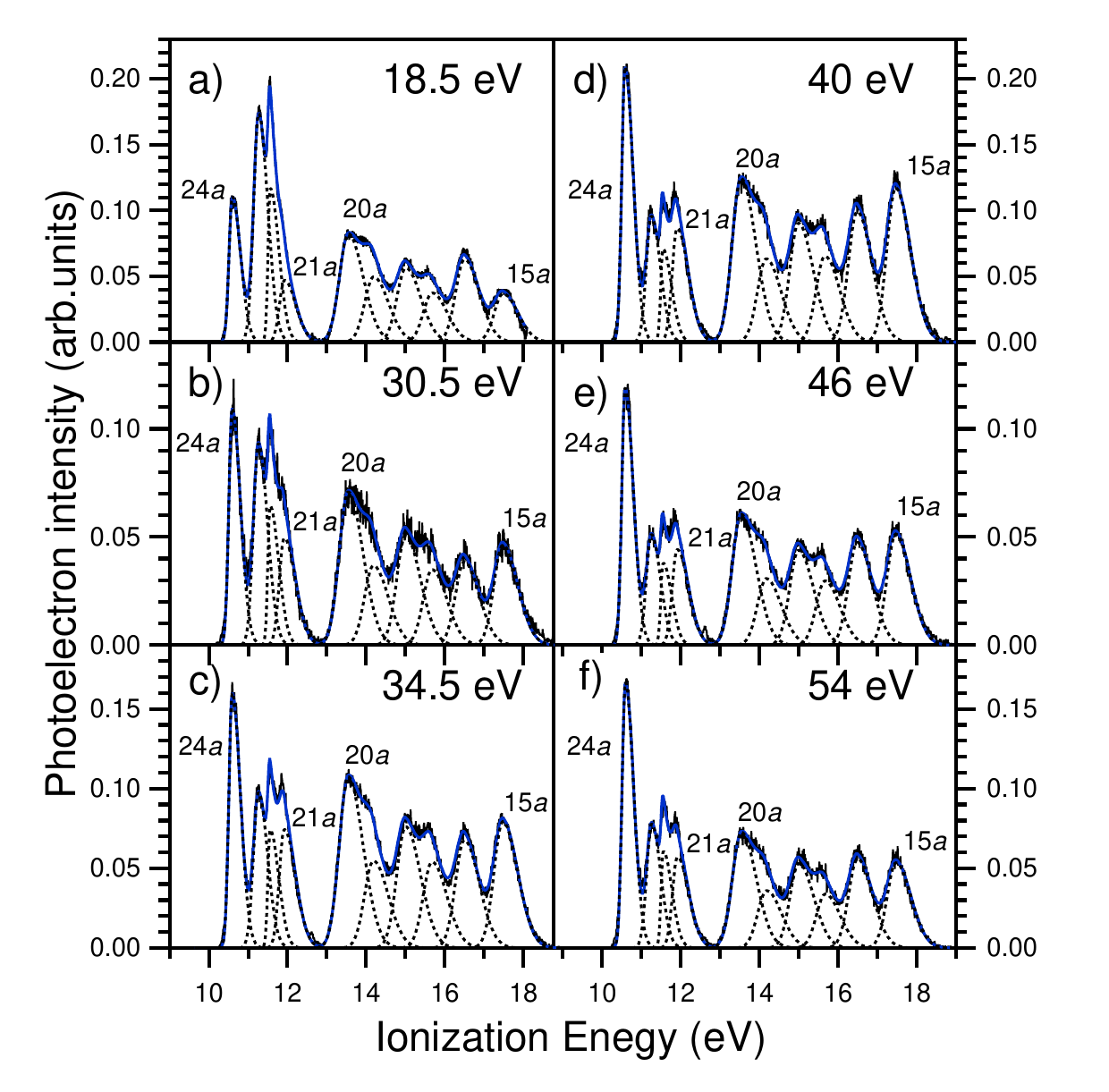}
    \caption{Outer valence PE spectra of epichlorohydrin recorded at the magic angle, 
    $\theta = 54.7^\circ$, and at different photon energies (panels a--f). 
    Spectral fit simulation curve (blue solid line) and bi-Gaussian fit components 
    (dotted lines), obtained by the global multipeak fit, are shown with the 
    experimental spectrum (solid thin line).}
    \label{fig:pe-Fitshape-epichlorohydrin}
\end{figure}

\section{Computational Details}
\label{sec:ComputationalDetails}
\begin{figure}[hbpt!]
    \centering
    \includegraphics[width=0.8\linewidth]{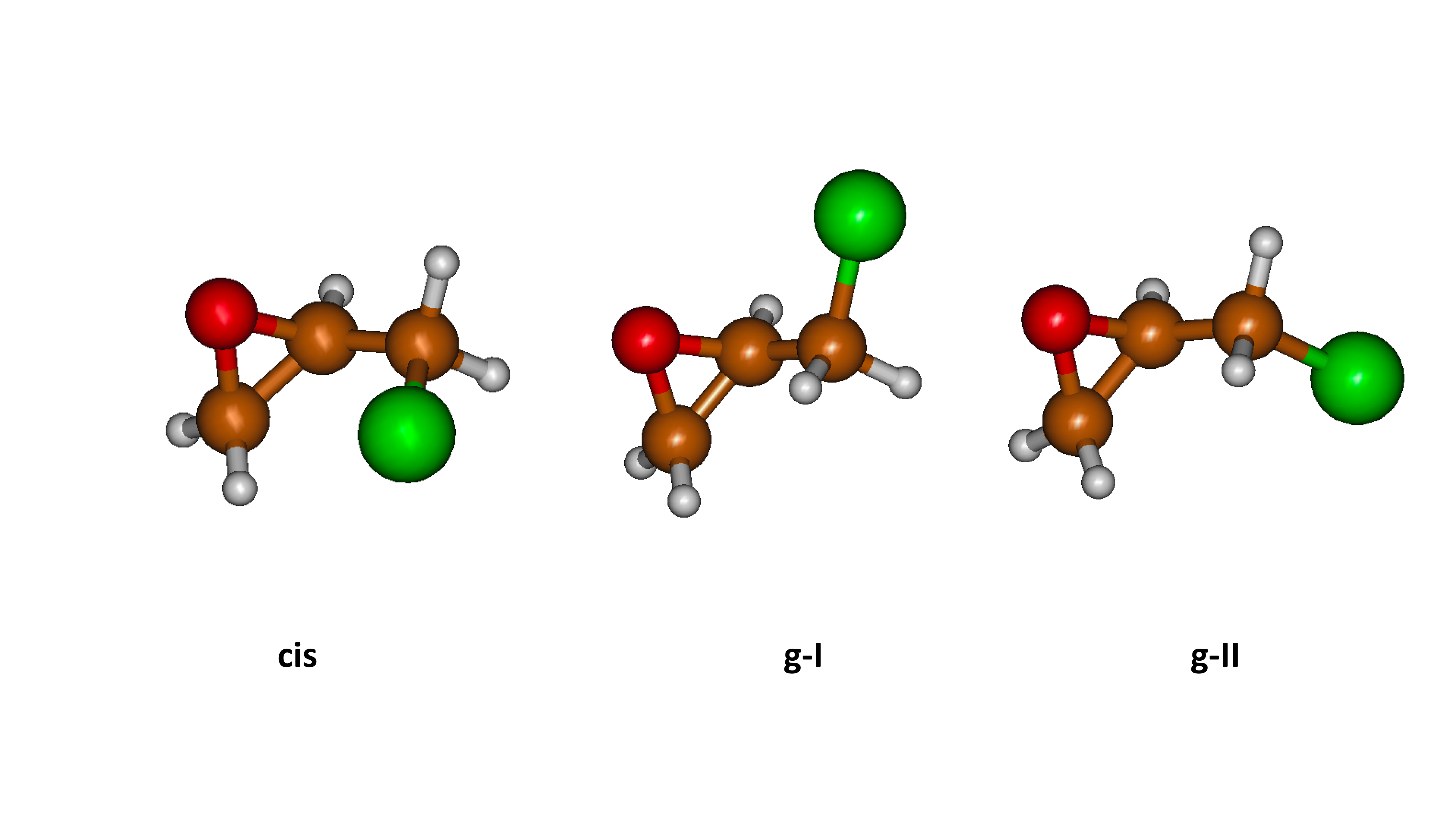}
    \caption{The three rotamers of epichlorohydrin.}
    \label{fig:rotamers}
\end{figure}

The three stable conformers cis, g-I and g-II \cite{Tam:2006:OR:epichlorohydrin,Stranges:2011} considered in this study are illustrated in Fig.~\ref{fig:rotamers}. They differ essentially by the rotation of the CH$_2$Cl group around the exocyclic C-C bond with respect to the oxirane ring. The corresponding Cartesian coordinates are reported in Section~\ref{sec:cartesian}.

The DFT and TDDFT calculations have been performed with the Tiresia program~\cite{Decleva:2022:Tiresia,Toffoli:2024:Tiresia} which employs a multicentric B-spline basis~\cite{Toffoli:2002:Bspline-multicenter,Bachau:2001:Bspline-review} 
which is able to accurately describe both bound and continuum orbitals. The algorithms have been presented in detail in previous publications.~\cite{Toffoli:2002:Bspline-multicenter,Decleva:2022:Tiresia,Toffoli:2024:Tiresia}
In the present case, the long-range 
one-center expansion has employed a linear knot sequence with range $R_{max}$ = 25. au and step size h = 0.50 au, and maximum angular momentum $L_{max}$ = 14. Short-range LCAO functions have employed $R_{maxi}$ ranging from 0.8-1.2 au for H and C atoms, 1.3 au for O and 2.0 for Cl, with step size about 0.2 au, $L_{maxi}$ = 1 for H and 2 for the heavier atoms. Tests have been performed to ensure that the results reported for $\sigma$ and $\beta$ are convergent. Initial density to define the GS Kohn-Sham Hamiltonian is obtained from a ground state calculation with the ADF program~\cite{ADF} 
employing the STO DZP basis from the ADF database.~\cite{ADF-STObasis} Both bound and continuum orbitals are then obtained with Tiresia as eigenvectors of the KS hamiltonian, employing the LB94 exchange-correlation potential.~\cite{vanLeeuwen:1994:LB94-photoioniz} 
In the fixed nuclei TDDFT calculations very sharp autoionization resonances appear at low energy, which are broadened by nuclear motion and are not seen in the experiment. To make comparison clearer the calculated profiles have been smoothed
{by a convolution with a binomial smoothing, also called gaussian smoothing with two passes}.~\cite{Smoothing} 

The ab initio EOM-CCSD calculations of the ionization energies and corresponding Dyson orbitals were carried out using \texttt{Q-Chem},~\cite{Qchem541} where the algorithms of Refs.~\citenum{Oana:2007,Oana:2009} are implemented.
The correlation consistent aug-cc-pVDZ basis set~\cite{Woon:1994} of Dunning and coworkers was used. The left and right Dyson orbitals were then fed into the Tiresia code,~\cite{Decleva:2022:Tiresia,Toffoli:2024:Tiresia} that had previously been adapted to the calculation of photoionisation dynamical parameters using the EOM-CCSD Dyson orbitals.~\cite{Moitra:2020}

\begin{figure}[hbpt!]
    \centering
\includegraphics[width=0.75\linewidth]{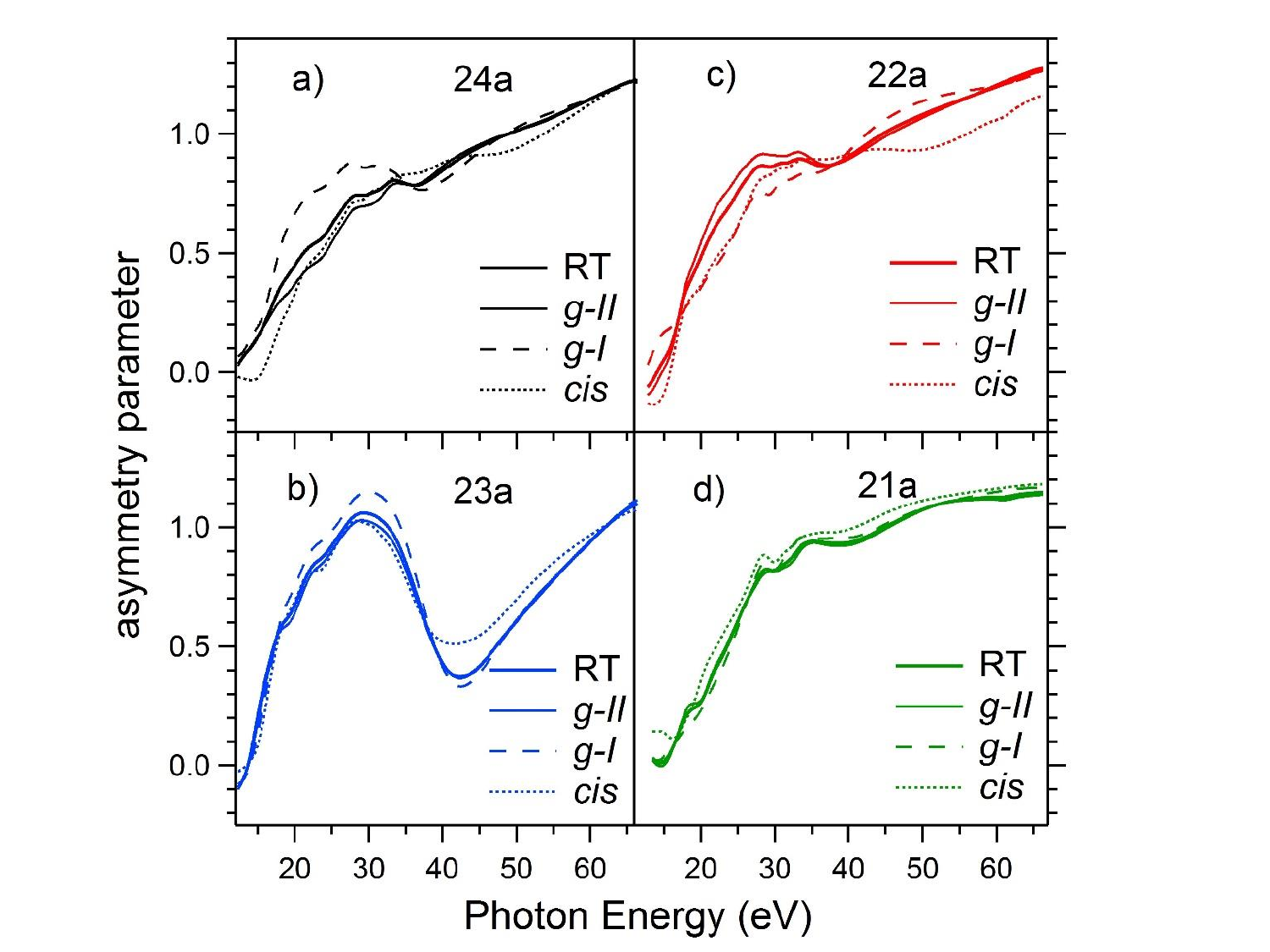}
    \caption{Epichlorohydrin. Anisotropy parameters of the stable rotamers, g-II (solid thin line), g-I (broken line), and cis (dotted line), for the outermost valence ionizations (panels a-d), calculated at {DFT-TDDFT} level. The room temperature Boltzmann average (curve RT, solid thick line) is also reported, along with the rotamers' anisotropy parameters.}
    \label{fig:beta_Rotamers_TDDFT}
\end{figure}

Relative energies for the three conformers have been evaluated previously~\cite{Stranges:2011}  as 0.00, 0.53 and 1.60 Kcal/mol for g-II, g-I and cis conformers, giving Boltzmann populations, at the experimental temperature  $T$=23~$^\circ$C, of 0.68, 0.28 and 0.04 respectively. 
An example of $\beta$ profiles for the three conformers and their average is displayed in Fig.~\ref{fig:beta_Rotamers_TDDFT}. 

Some differences are apparent, especially for the cis conformer, whose population is, however, very small, and the average is close to the g-II profile. All $\sigma$ and $\beta$ profiles reported in the paper are corresponding averages over the conformers.  

\section{Higher energy ionizations (deeper ionisations)}
\label{sec:otherchannels}

Experimental and theoretical results have also been obtained for the remaining part of the outer valence shell, 20a-15a ionizations (an expanded experimental photoelectron spectrum is presented in Ref.~\citenum{Stranges:2011}). Together with the experimental values, we present theoretical results relative only to the most accurate TDDFT continuum. For the deepest 15a ionization only initial state DFT results are evaluated, while all DFT, HF and Dyson initial orbital are presented for the other ones.

\begin{figure}[hbpt!]
    \centering
\includegraphics[width=0.75\linewidth]{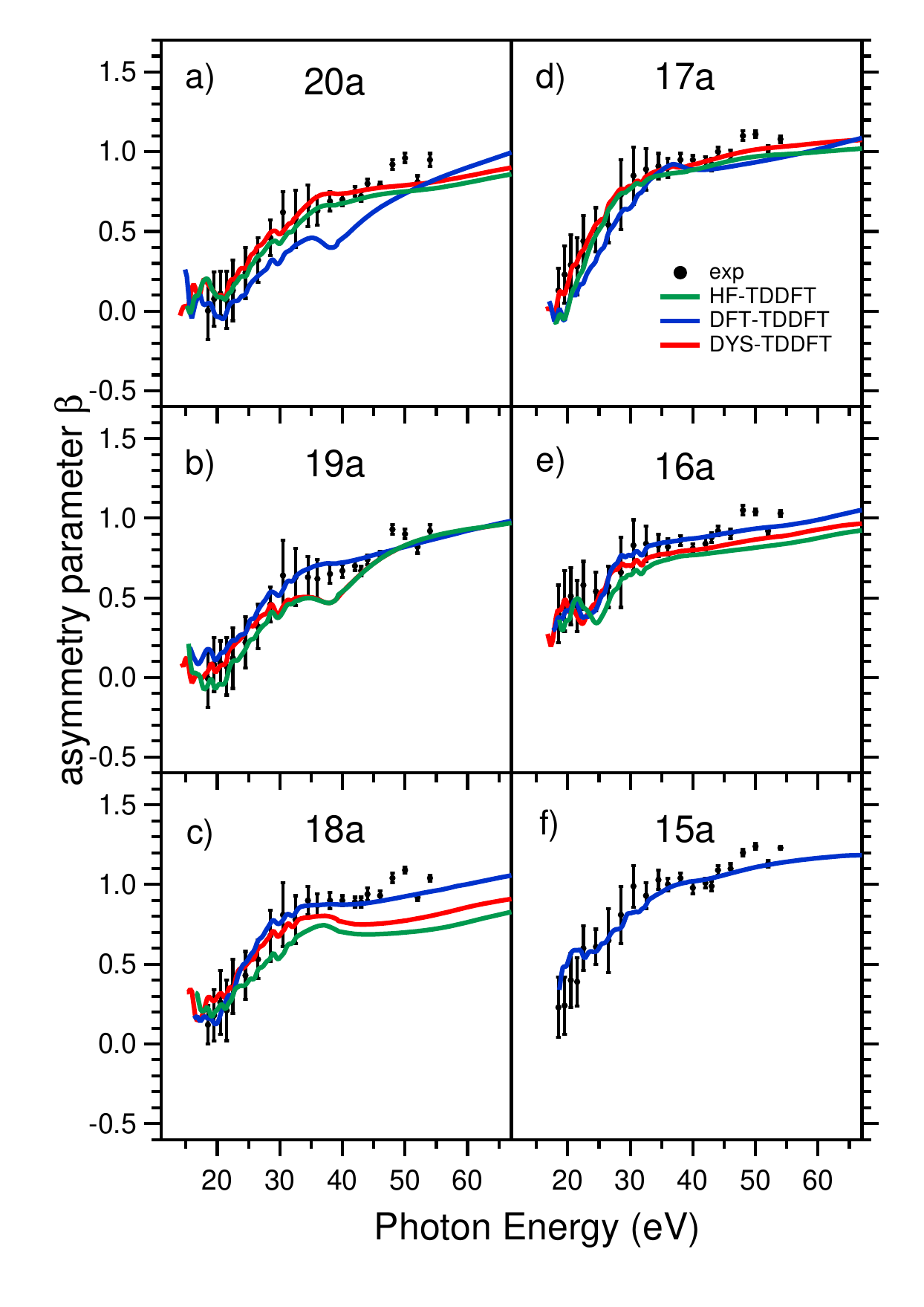}
    \caption{Epichlorohydrin. Asymmetry parameters $\beta$ as a function of the photon energy for orbital 20a, 19a, 18a, 17a, 16a and 15a. Experiment versus computational results obtained using a TDDFT representation of the continuum together with Hartree-Fock (label HF-TDDFT), Density Function Theory (label DFT-TDDFT) and EOM-CCSD  Dyson (label DYS-TDDFT) orbitals.}
    \label{fig:beta:20a-15a}
\end{figure}

The results for the $\beta$ parameter are reported in Figure~\ref{fig:beta:20a-15a}. 

Looking at the experimental results, and comparing with the four outermost ionizations, a few points can be observed. At low energy, the $\beta$ value increases smoothly, less steeply than in the outermost ones, the only exception being the 16a ionization, which displays a steep increase and a sudden flattening. Then at high energy all profiles seem to flatten around $\beta=1$ value. Such behaviour is typical of $\sigma$ orbitals of organic moieties. There are also some non conspicuos features in specific ionizations, like a kind of shallow depression around 40 eV (perhaps a sign of the CM minimum through interchannel coupling) in several channels and a dip at low energy, about 25 eV, in 16a. Generally a very satisfactory agreement with Dyson-TDDFT results is observed. Only the dip in 19a is overestimated, and too low $\beta$ is displayed at high energy in 18a. More remarkable is the good general agreement of HF results with the Dyson ones, not surprising given the near validity of Koopmans theorem obtained in this region, with Dyson pretty close to HF orbitals. On the other hand, somewhat larger differences are afforded by the DFT MOs, notably in the 18a-20a channels. Similar behaviour has been observed before~\cite{Moitra:2021} and may be due to a peculiarity of the LB94 $V_\textrm{XC}$ potential currently employed, or a feature of the DFT approach. This would be worth further investigation. 

\begin{figure}[hbpt!]
    \centering
\includegraphics[width=0.75\linewidth]{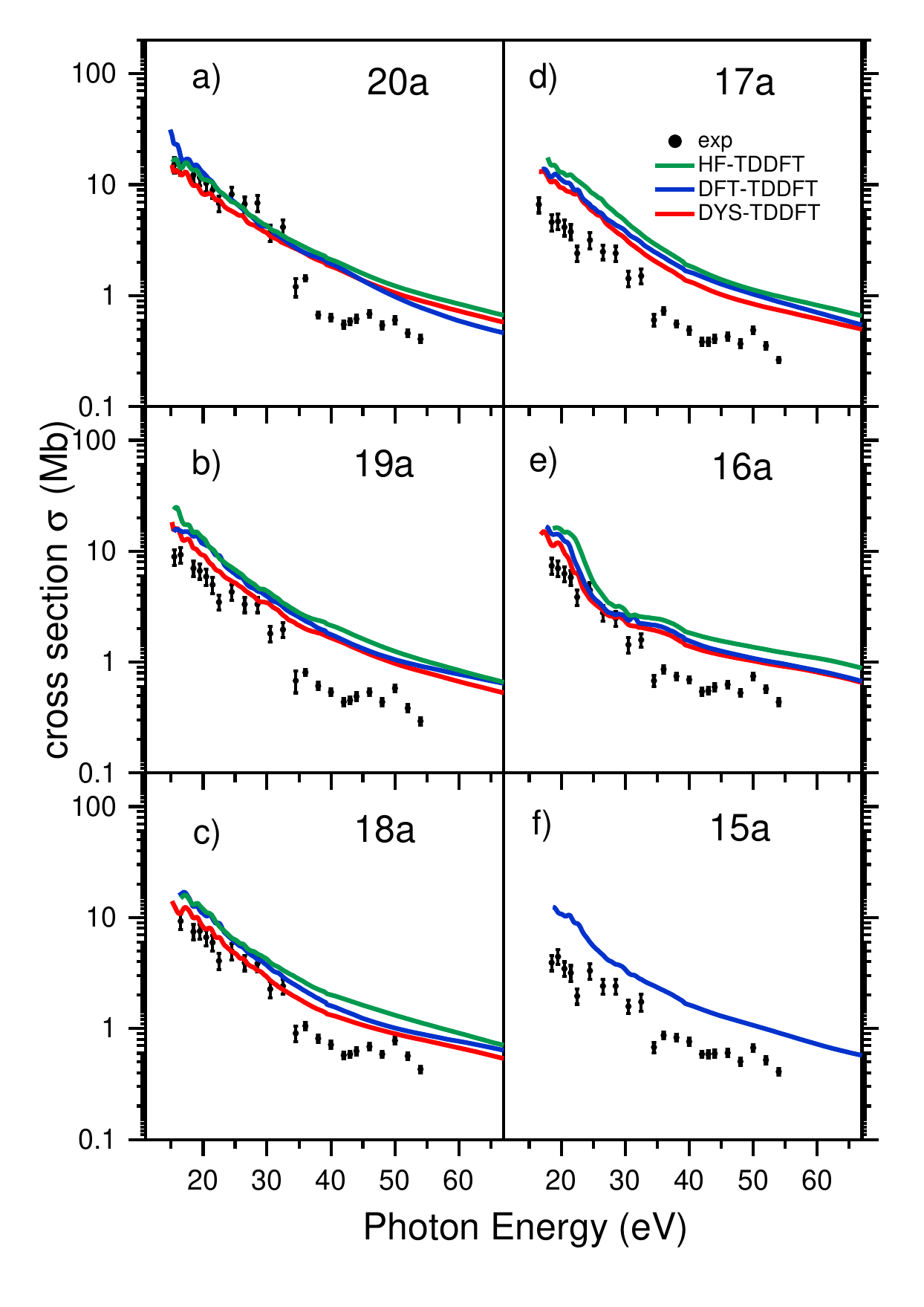}
    \caption{Epichlorohydrin. Cross-section $\sigma$ as a function of the photon energy for orbitals 20a, 19a, 18a, 17a, 16a and 15a. Experiment versus computational results obtained using a TDDFT representation of the continuum together with Hartree-Fock (label HF-TDDFT), Density Function Theory (label DFT-TDDFT) and EOM-CCSD Dyson (label DYS-TDDFT) orbitals.}
    \label{fig:sigma:20a-15a}
\end{figure}

\begin{figure}[hbpt!]
    \centering
\includegraphics[width=0.75\linewidth]{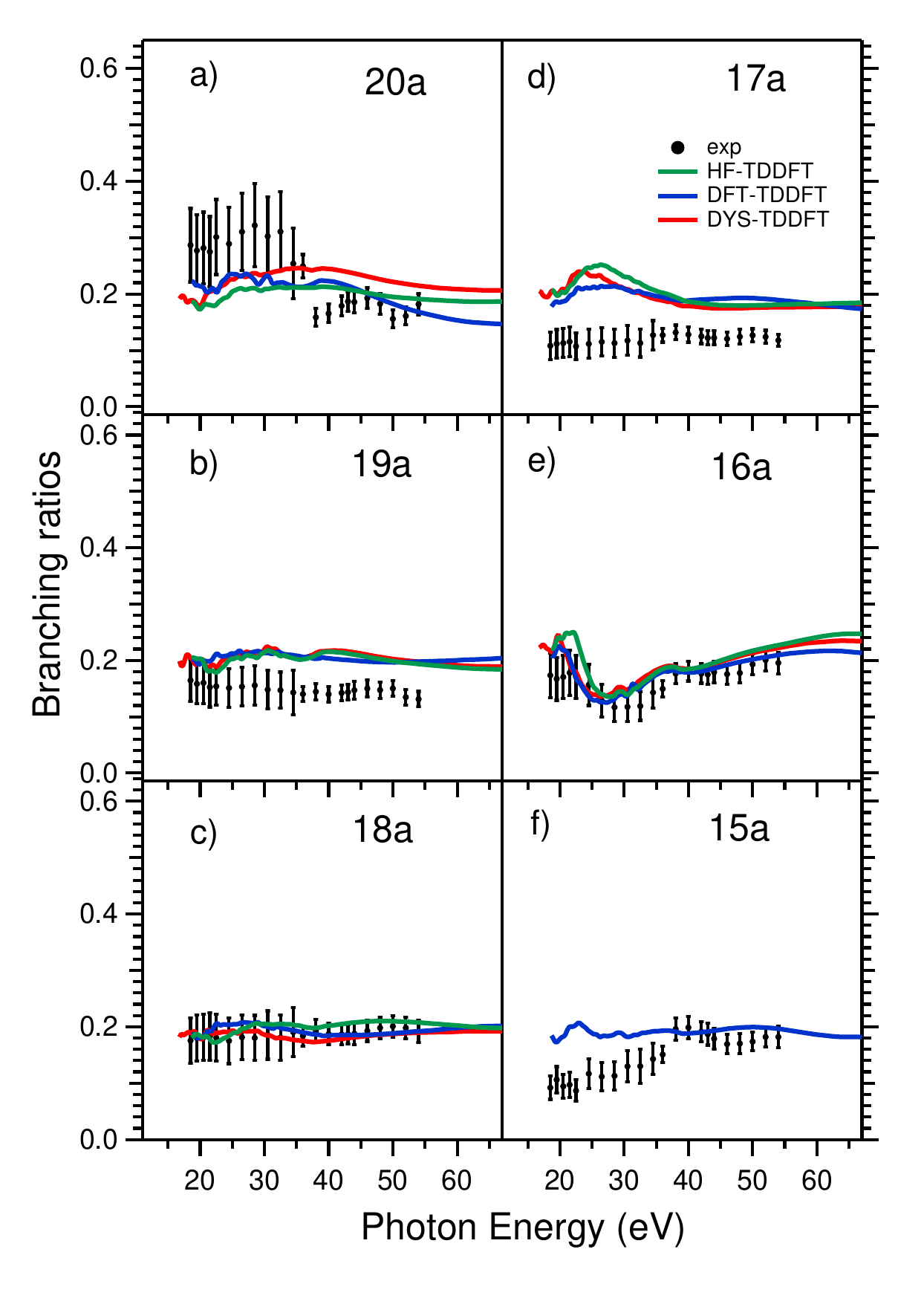}
    \caption{Epichlorohydrin. Branching ratios as a function of the photon energy for orbitals 20a, 19a, 18a, 17a, 16a and 15a. Experiment versus computational results obtained using a TDDFT representation of the continuum together with Hartree-Fock (label HF-TDDFT), Density Function Theory (label DFT-TDDFT) and EOM-CCSD Dyson (label DYS-TDDFT) orbitals.}
    \label{fig:Branching:20a-15a}
\end{figure}

The cross sections, reported in Fig.~\ref{fig:sigma:20a-15a}, are computed to be very similar and pretty flat, with the exception of some structure present in the 16a ionization. 
Also, all initial orbitals give pretty similar results, with good agreement among themselves. 
On the contrary, more structure is seen in the experimental results, and in particular a sudden drop of $\sigma$ in the 30-40 eV region, reminiscent of the CM structure, that could be propagated through interchannel coupling effects, since negligible Cl orbital participation is expected in this region (see MO composition), despite the fact that IC is in principle included in the TDDFT description. 
So, agreement with the calculated values is not good, the experimental values being consistently lower above about 35eV.  
This behaviour largely cancels out in the BR plots displayed in Fig.~\ref{fig:Branching:20a-15a} but still showing notable disagreements. 
Experimental results for 19a, 18a and 17a show a very flat BR. 
The agreement with the theoretical values is excellent for 18a, still good for 19a, although slightly overestimated by theory, while a low energy maximum predicted by theory is not observed in 17a, where the calculated values are also significantly overestimated.  Some structure is observed in the other BRs. In the 16a channel, a low energy dip is very well matched by the theoretical results, with excellent overall agreement. Flatter BRs are theoretically predicted for the remaining 20a and 15a ionizations, while more structure in the experimental results is not reproduced, although the average value is quite well matched.
As noted for the outermost ionizations, accurate reproduction of the cross sections is more difficult, as it is the accurate experimental fitting of the corresponding bands, so we cannot unambiguosly attribute the observed disagreements to deficiencies in either the theoretical or experimental values, or possibly both. It is fair, however, to note that a pretty satisfactory overall description is obtained by the present EOM-CCSD Dyson-TDDFT approach.

\clearpage

\section{Cartesian coordinates of the three rotamers (Bohr units)}
\label{sec:cartesian}

\begin{verbatim}
Cis 
C    0.614341   -1.086131   -2.735336
C    1.353776   -2.325133   -0.396607
O    2.708330   -0.205827   -1.289086
C   -0.008141   -2.034237    2.038108
Cl  -2.006889    0.656667   2.140174
H   -1.044695    0.155827   -2.704761
H    1.101202   -1.974655   -4.544451
H    2.386020   -4.130124   -0.528796
H    1.369614   -1.817368    3.581519
H   -1.181775   -3.709235    2.413898
\end{verbatim}

\begin{verbatim}
g-I 
C   -2.707614    0.570722   -2.570644
C   -0.590661   -0.651584   -1.291752
O   -0.306921    1.750101   -2.433162
C   -0.485396   -0.682785    1.493644
Cl   2.707694   -0.570690    2.570566
H   -4.197911    1.452797   -1.427147
H   -3.275343   -0.096787   -4.448736
H    0.412040   -2.193443   -2.258217
H   -1.333144   -2.417852    2.267780
H   -1.459268    0.987444    2.261888
\end{verbatim}

\begin{verbatim}
g-II 
C   -1.730567    1.567111   -2.674188
C    0.362412    0.258683   -1.464677
O    0.702434    2.702350   -2.507699
C    0.552840    0.187234    1.325315
Cl  -0.702291   -2.702403    2.507681
H   -3.191889    2.419594   -1.473982
H   -2.313712    1.013552   -4.583473
H    1.301116   -1.291061   -2.481441
H    2.534850    0.306269    1.942803
H   -0.517135    1.764267    2.158405
\end{verbatim}

\clearpage

\section{Additional Tables and Figures}

% \begin{table}[htb!]
% \centering
% \caption{Epichlorohydrin. Ionization energies (IE, eV) and pole strengths $R_F$ computed using EOM-CCSD/aug-cc-pVDZ for the three conformers.}
% \label{tab:IEs:conformers}
% \begin{tabular}{c|cc|cc|cc}
% \hline
%   & \multicolumn{2}{c|}{cis} & \multicolumn{2}{c|}{g-I} & \multicolumn{2}{c}{g-II}   \\
% Peak $i$ & IE (eV) & $R_F$ &IE (eV) & $R_i$ &IE (eV) & $R_i$  \\\hline
% 1/A & 10.2871 & 0.8917 & 10.4807 &  0.8917  & 10.5111  & 0.8906 \\  
% 2/A & 10.9981 & 0.9082 & 11.0268 &  0.9062  & 11.1611  & 0.9096 \\  
% 3/A & 11.3221 & 0.9052 & 11.1343 &  0.9056  & 11.4109  & 0.9051  \\ 4/A & 11.6759 & 0.9005 & 11.7241 &  0.9013  & 11.6467  & 0.9005  \\ 5/A & 13.4933 & 0.9046 & 13.7858 &  0.8983  & 13.6111  & 0.8970  \\ 6/A & 13.7972 & 0.8922 & 14.0289 &  0.8980  & 13.9861  & 0.9045  \\ 7/A & 14.6238 & 0.8999 & 14.5485 &  0.8977  & 15.1296  & 0.8946  \\ 8/A & 16.0641 & 0.8935 & 15.7938 &  0.8952  & 15.4870  & 0.8941  \\ 9/A & 16.4279 & 0.8876 & 16.6666 &  0.8861  & 16.5929  & 0.8861  \\ 10/A & 17.684  &  0.8791 & 17.6781  &  0.8805  & 17.6996  & 0.8821  \\  
% \hline
% \end{tabular}
% \end{table}

\begin{table}[htb!]
\centering
\caption{Epichlorohydrin. 
HF/aug-cc-pVDZ and DFT (LB94/DZP) molecular orbital energies (eV).}
\label{tab:hf-dft-mo}
    \begin{tabular}{c|cc|cc|cc}
    \hline
         & \multicolumn{2}{c|}{cis} & \multicolumn{2}{c|}{g-I} & \multicolumn{2}{c}{g-II}   \\
     MO  & HF & DFT & HF & DFT & HF & DFT \\
         \hline
         24a & 11.728 & 11.927 & 11.810 & 12.061 & 11.973 & 12.142\\
         23a & 11.945 & 12.177 & 11.891 & 12.137 & 12.054 & 12.279\\
         22a & 12.353 & 12.567 & 12.381 & 12.461 & 12.517 & 12.571\\
         21a & 12.898 & 13.232 & 12.843 & 13.334 & 12.816 & 13.314\\
         20a & 14.422 & 14.470 & 14.884 & 14.721 & 14.585 & 14.738\\
         19a & 15.075 & 15.152 & 15.211 & 15.296 & 15.102 & 15.073\\
         18a & 15.673 & 15.662 & 15.564 & 15.669 & 16.272 & 16.118\\
         17a & 17.442 & 16.649 & 17.116 & 16.553 & 16.871 & 16.362\\
         16a & 18.095 & 17.224 & 18.394 & 17.441 & 18.340 & 17.380\\\hline
    \end{tabular}
\end{table}

\clearpage

\begin{table}[htb!]
\centering
\caption{Mulliken 
atomic 2p (and Cl 3p) populations of the HF MO.
C1 is the -CH$_2$ carbon in the ring, C2 is the CH carbon and C3 is the carbon bound to the Cl atom.
Other populations, with the exception of H 1s, are very small and therefore omitted.
}
\label{tab:Mulliken}
\begin{tabular}{lccccccccccc}\\
\hline\hline
\multicolumn{12}{c}{g-II}\\\hline
     &  & \multicolumn{10}{c}{MO} \\\cline{3-12}
Atom &  Orb.
& 15 & 16 & 17  &   18  &  19 & 20  &    21   &   22    &  23  &  24 \\
\hline\hline
C 1 &  p  &  0.263 &  0.340 &  0.054  & 0.154 &  0.325  & 0.109 &  0.127 &  0.484 &  0.010 &  0.134 \\
C 2  & p  & 0.295 & 0.284 &  0.070 &  0.088 &  0.321 &  0.098 &  0.093 &  0.415  & 0.089 &  0.105
\\
O  &   p &   0.422 &  0.441  & 0.201 &  0.635   & 0.006  & 0.536 &  1.150 &  0.377 &  0.281 &  0.097 \\
C 3 &  p &   0.223 &  0.419 &  0.744 &  0.374   & 0.336  & 0.314 &  0.016 &   0.028 &  0.081   & 0.103 \\
Cl  &  p  &  0.038 &  0.057 &  0.226 &  0.355  & 0.164  & 0.586 &  0.322 &  0.539 &  1.428 &  1.281 \\\hline\hline
\multicolumn{12}{c}{g-I}\\\hline
C 1 &  p  &   0.261 &  0.364 &  0.199  & 0.168   & 0.138 &  0.227 &  0.140 &  0.393  & 0.000  & 0.169 \\
C 2 &  p  &   0.295 &  0.295 &  0.107 &  0.035 &   0.232  & 0.238  & 0.111 &  0.366 &  0.009 &  0.195 \\
O  &   p &    0.472 &  0.413  & 0.207 &  0.271 &   0.479 &  0.406  & 1.033  & 0.638 &  0.044 &  0.160 \\
C 3 &  p  &   0.238 &  0.402 &  0.647 &  0.581 &  0.419 &  0.174 &  0.025 &  0.050 &  0.080 &  0.065 \\
Cl &   p  &   0.037 &  0.048 & 0.161 &  0.522 &   0.371  & 0.117 &  0.421 &  0.263 &  1.702 &  1.340 \\
\hline\hline
\multicolumn{12}{c}{cis}\\\hline
C 1 &  p  &   0.250 &  0.335 &   0.017 &  0.217  & 0.240  & 0.162 &  0.141 &  0.432 &  0.053  & 0.125  \\
C 2 &  p  &   0.293 &  0.325 &  0.101 &  0.088 &  0.152 &  0.224   & 0.115  & 0.365 &  0.113 &  0.094 \\
O & p & 0.460 &  0.453 &  0.097  & 0.196 &  0.849 &  0.137 &  0.923  & 0.487 & 0.294 &  0.221 \\
C 3  & p &  0.273 &  0.284 &   0.887 &  0.522 &  0.293 &  0.318  & 0.018 &  0.021 &  0.082  & 0.098 \\
Cl  &  p  &   0.056 &  0.085 &  0.156 &  0.459 &  0.089 &  0.542  & 0.608  & 0.499 &  1.265  & 1.229\\\hline
\end{tabular}
\end{table}

\begin{figure}[hbpt!]
    \centering
\includegraphics[width=0.7\linewidth]{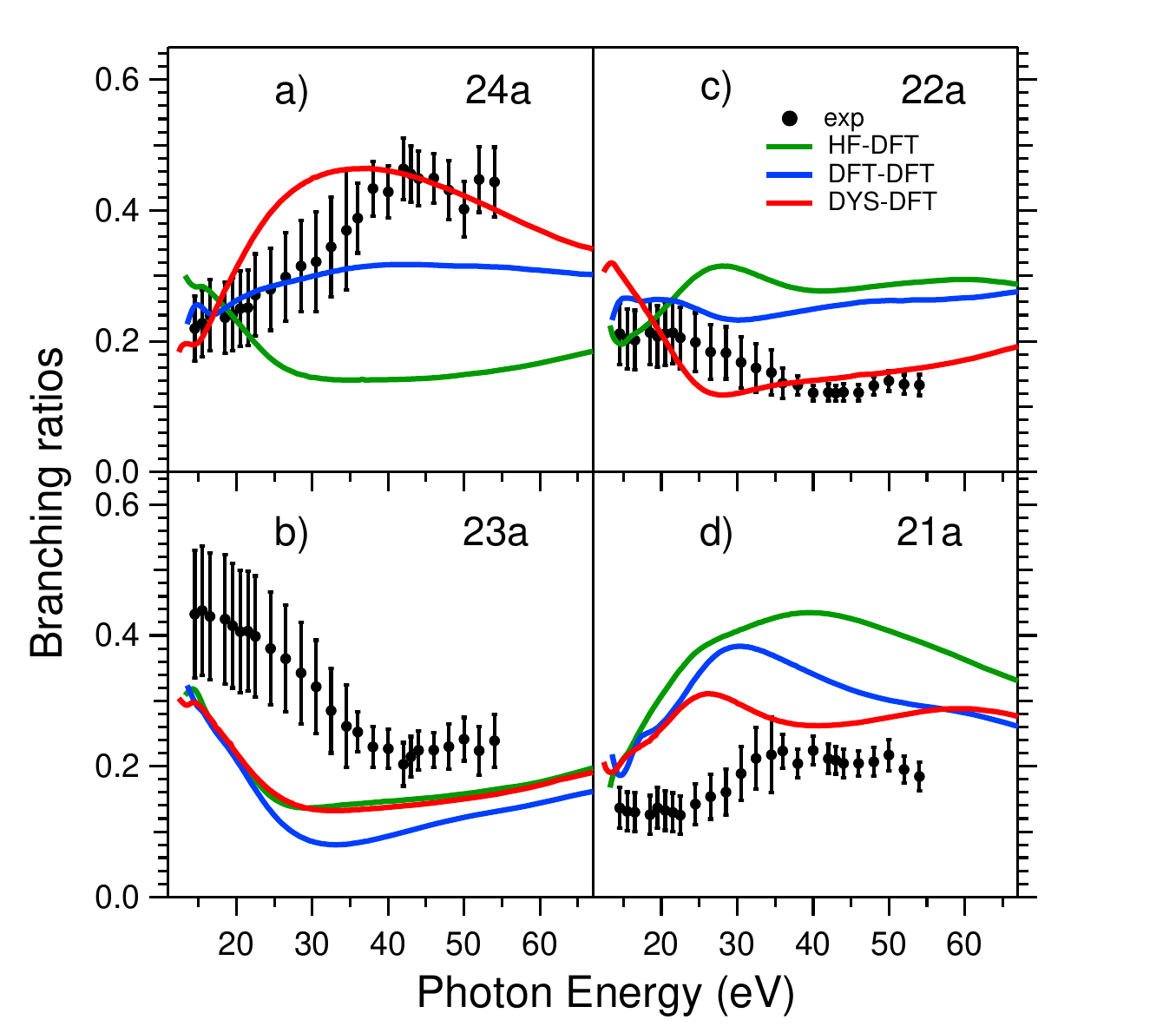}
\caption{Epichlorohydrin. Branching ratios for orbitals 24, 23, 22 and 21 as a function of the
photon energy. Experiment versus computational results obtained using
a DFT representation of the continuum together with Hartree-Fock
(HF-DFT), Density Function Theory (DFT-DFT) and EOM-CCSD Dyson (DYS-DFT) orbitals}
    \label{fig:BR:24-21:DFT}
\end{figure}

\clearpage
\section{Comparison between experimental and theoretical observables}

The sample experimentally investigated is modeled as a gaseous mixture of the three stable rotamers of epichlorohydrin, present with their Boltzmann relative weights at the experimental temperature, namely g-II (0.68), g-I (0.28), and cis (0.04). 
The theoretical calculations provide the photoionization observables for each of the three rotamers, thus allowing to derive the average value of the calculated observable, which is compared to the experimental one.

\paragraph{Partial cross sections} In addition to what described in the main text, only additional complementary details are given here. The experimental partial cross sections for the 20a-15a ionization channels are compared to the corresponding calculated values in Figure~\ref{fig:sigma:20a-15a} using the same conversion factor derived for the comparison of the 24a-21a channels shown in Fig. 4 and Fig. 5. The branching ratio of the 20a-15a ionizations reported in Figure~\ref{fig:Branching:20a-15a} is given as the ratio between the PE peak area of the specific ionization and the sum of the corresponding areas of the 20a-15a ionizations.

\paragraph{Asymmetry parameters} All the measured experimental asymmetry parameters $\beta_{\textrm{ex}}^j$, in 
Fig.~2, Fig.~3, and Fig.~\ref{fig:beta:20a-15a}, are compared to the corresponding average calculated value $\beta_{\textrm{av}}^j$, which is derived from the following formula,
\begin{equation}
    \beta_{i,\textrm{av}} =
    \frac{\sum_j p_j \sigma_{i,j} \beta_{i,j}}{\sum_j p_j \sigma_{i,j}}
\end{equation}
where $p_j$ is the Boltzmann relative weight of the $j$-th rotamer (with $j$ = g-II, g-I, and cis), and $i$ is referring to the $i$-th ionization channel.
The formula for the average is derived in the next section.

\section{Derivation of $\beta_{\textrm{av}}$ over conformers}

Let us consider a statistical mixture of $N$ conformers with populations (Boltzmann weights)
\begin{equation} p_j, \quad\quad \sum_j p_j = 1 \quad\quad j=1,\ldots,N 
\end{equation}
Omitting the subscript $i$ for the ionisation channel for ease of notation, the differential cross section for each conformer  is given by
\begin{equation} 
\label{CS1}
\frac{d\sigma_{j}}{d\theta} = \frac{\sigma_j}{4\pi} \; [ 1 + \beta_j P_2(\cos \theta) ]  \quad j = 1,\ldots,N
\end{equation}
which is valid at all angles.
The total differential cross section is then given by the weighted sum of the individual terms
\begin{equation} 
\label{CS2}
\frac{d\sigma_{\textrm{av}}}{d\theta} = \sum_j p_j \frac{d\sigma_{j}}{d\theta}~.
\end{equation}
Substituting eq.~\eqref{CS1} in the right hand side one obtains
\begin{eqnarray}
 \nonumber
  \frac{d\sigma_{\textrm{av}}}{d\theta} &=& \sum_j \; p_j  \frac{\sigma_j}{4\pi} \; [ 1 + \beta_j P_2(\cos \theta) ] 
  \\\nonumber
   &=&  \frac{1}{4\pi} \left( \sum_j \; p_j \sigma_j +  \sum_j \; p_j \sigma_j \beta_j P_2(\cos \theta)  \right)
   \\\nonumber
   &=&  \frac{1}{4\pi} 
   \left\{ 
   \sum_j \; p_j \sigma_j +  \left(\sum_j \; p_j \sigma_j \beta_j\right) 
   P_2(\cos \theta) 
   \right\} 
   \\\nonumber
   &=&  \frac{1}{4\pi} 
   \left( \sigma_{\textrm{av}} +  B_{\textrm{av}} P_2(\cos \theta) \right) 
   \\
   &=&  \frac{\sigma_{\textrm{av}}}{4\pi} \; [1  +  \beta_{\textrm{av}} P_2(\cos \theta) ]
\end{eqnarray}
with
\begin{equation} \label{CS3}
 \sigma_{\textrm{av}} = \sum_j p_j \sigma_j \quad \quad \beta_{\textrm{av}} = \frac{B_{\textrm{av}}}{\sigma_{\textrm{av}}} = \frac{\sum_j  p_j \sigma_j \beta_j}{\sum_j p_j \sigma_j}
\end{equation}
which is still valid at all angles. 
Thus, if one measures the average differential cross section at two angles (e.g., zero and the magic angle),
one can extract from this the value of $\beta_{\textrm{av}}$, obtained by averaging the individual $\beta_j$ weighted as in eq~\eqref{CS3}.

\clearpage
\bibliographystyle{rsc}
\bibliography{rsc,stefano}